\documentclass{elsarticle}
\usepackage{graphicx}
\usepackage{amsmath,amsthm,amssymb,amsfonts,mathtools,bbold,hyperref}
\usepackage{natbib,geometry,fleqn}
\usepackage{cite}

\DeclareMathOperator{\tr}{tr}

\newcommand{\A}{\mathcal{A}}

\begin{document}

\title{Entanglement entropy and non-local duality: quantum channels and quantum algebras}

\author[1]{Oliver DeWolfe}
\ead{oliver.dewolfe@colorado.edu}

\author[1]{Kenneth Higginbotham}
\ead{kenneth.higginbotham@colorado.edu}

\affiliation[1]{organization={Department of Physics and Center for        Theory of Quantum Matter, University of Colorado Boulder},
    addressline={390 UCB},
    city={Boulder, CO},
    postcode={80309},
    country={USA}}

\begin{abstract}
We investigate the transformation of entanglement entropy under dualities, using the Kramers-Wannier duality present in the transverse field Ising model as our example. Entanglement entropy between local spin degrees of freedom is not generically preserved by the duality; instead, entangled states may be mapped to states with no local entanglement. To understand the fate of this entanglement, we consider two quantitative descriptions of degrees of freedom and their transformation under duality. The first involves Kraus operators implementing the partial trace as a quantum channel, while the second utilizes the algebraic approach to quantum mechanics, where degrees of freedom are encoded in subalgebras. Using both approaches, we show that entanglement of local degrees of freedom is not lost; instead it is transferred to non-local degrees of freedom by the duality transformation.
\end{abstract}

\maketitle

\section{Introduction}

The importance of dualities cannot be overstated. They have become a favored tool in many areas of current research,  notable for their ability to turn intractable problems into manageable solutions. In recent years, the field of quantum information theory has begun to contribute to the study of dualities, particularly through the AdS/CFT correspondence. Breakthroughs such as the holographic realization of entanglement entropy \citep{PhysRevLett.96.181602,Hubeny:2007xt,Faulkner:2013ana,Engelhardt:2014gca}, entanglement wedge reconstruction \citep{Czech:2012bh, Wall:2012uf, Headrick:2014cta,Cotler:2017erl}, and quantum error correction \citep{Almheiri:2014lwa,Pastawski:2015qua,Dong:2016eik,Harlow:2016vwg} have been invaluable to the study of the AdS/CFT correspondence and beyond. Indeed, the nature of quantum gravity seems intimately connected to how quantum information is expressed through dualities.

An essential theme in the applications of quantum information is entanglement. Our focus in this work will be on the interplay between entanglement and duality. In general, duality transformations act nontrivially on the entanglement of a quantum system. A duality rearranges the system's degrees of freedom into a new set of degrees of freedom, in general non-locally related to the original. However, it is well understood that expressing a state in a different set of degrees of freedom does not necessarily preserve its entanglement \citep{hategan_entanglement_2018, de_la_torre_entanglement_2010, zanardi_quantum_2004, harshman_observables_2011, jakubczyk_alternative_2015}. Thinking of a duality as a change of degrees of freedom, it should thus affect the entanglement entropy of a state in a non-trivial way, in general turning an unentangled state into an entangled state and vice versa.

It is then natural to ask, when duality turns an entangled state into an unentangled state, whether the entanglement is lost, or is still present in some sense. We suggest that the right way to think about this is that while the entanglement of local degrees of freedom is not preserved, it is still present among a hidden set of non-local degrees of freedom. The main purpose of this paper is to describe ways to characterize these non-local degrees of freedom, and then to demonstrate that under duality entanglement is indeed passed from local to non-local variables.

Spin models provide an excellent starting ground for our considerations, as they contain a multitude of well understood dualities (see for example \citep{savit_duality_1980, wegner_duality_2014}). We will employ the transverse field Ising model (TFIM), which possesses the Kramers-Wannier duality, under which local spin degrees of freedom are carried to non-local combinations of spins characterizing frustration between neighbors. These ``domain walls" may be treated as new local degrees of freedom on a dual lattice. With suitable boundary conditions, the transverse field Ising model is exactly self-dual. These boundary conditions allow for a particularly simple implementation of Kramers-Wannier, where the duality acts as a unitary operator on the Hilbert space, mapping the Hilbert space to itself under a rotation and producing new degrees of freedom that are non-locally related to the original spins. As we shall discuss, it is easy to see that local entanglement of spins is not preserved under this duality.

To understand how the entanglement is passed into non-local variables, we must have a way to characterize a non-local degree of freedom. We consider two complementary approaches. For the first, a degree of freedom may be understood in terms of the operation of the partial trace, which removes one or more degrees of freedom while holding others invariant. For local degrees of freedom this is straightforward, but we describe how to generalize the idea of partial trace to non-local variables as well. To do so, it is useful to characterize the partial trace in the language of quantum channels, where tracing out the degree of freedom is implemented by the action of certain Kraus operators. For the second approach, we make use of the algebraic formulation of quantum mechanics, where the quantum system is characterized by an algebra instead of a Hilbert space, and a degree of freedom corresponds to a subalgebra. In both cases, we show how duality acts on the degrees of freedom, and demonstrate how a state with no local entanglement may be shown to possess entanglement between the non-local degrees of freedom. The two approaches are largely independent and may be used separately to understand the transformation of entanglement under duality.

Sec.~\ref{sec:Ising} begins with a review of the transverse field Ising model and Kramers-Wannier duality. We discuss open, periodic, and self-dual boundary conditions, and demonstrate how duality relates entangled to (locally) unentangled states. Sec.~\ref{sec:QChannels} reviews quantum channels and describes how the partial trace characterizes a degree of freedom by a set of Kraus operators for a quantum channel, and uses this approach to calculate the entanglement entropy of non-local degrees of freedom, showing that in this sense entanglement is preserved by duality. Sec.~\ref{sec:Algebras} begins with a review of algebraic quantum mechanics and the GNS construction used to obtain a Hilbert space from the defining algebra. The algebraic description of a degree of freedom as a subalgebra and the process for calculating entanglement entropy are then given, and these tools are used to verify the results of Sec.~\ref{sec:QChannels}, again discovering the entanglement of non-local degrees of freedom.

Related questions regarding the transformation of entanglement under electromagnetic duality in continuum field theories have been considered by \citep{donnelly_electromagnetic_2017,moitra_entanglement_2019,headrick_bosefermi_2013}. Previous work on the transformation of entanglement under dualities in lattice models, including the Kramers-Wannier duality, has been performed in \citep{Radicevic:2016tlt} using a primarily algebraic approach. Sec.~\ref{sec:Conclusions} offers a comparison with this previous investigation and concludes the present work.

\section{The transverse field Ising model and local entanglement}
\label{sec:Ising}

We begin with a review of the transverse field Ising model (TFIM). Consider a one-dimensional chain of $N$ sites, each site containing a spin-1/2 (qubit). The Hamiltonian for the TFIM is constructed from pairs of $Z$ Pauli operators to enforce nearest neighbor couplings and single site Pauli $X$ operators acting as the transverse field:
\begin{equation} \label{eq:ham_TFIM_open}
    H_{\text{TFIM}} = -J_z\sum_{n=1}^{N-1} Z_n Z_{n+1}- J_x\sum_{n=1}^N  X_n \,,
\end{equation}
where two couplings $J_x$ and $J_z$ control the strength of these terms. The dependence of this system's phases on these couplings is well understood. For $\lambda \equiv J_z/J_x \gg 1$, the nearest neighbor coupling dominates, and energy is minimized when all sites are aligned in the same $z$-direction. Since no absolute direction is preferred, when $J_x = 0$ strictly there are two degenerate ground states: $|\ldots\uparrow\uparrow\ldots\rangle$ and $|\ldots\downarrow\downarrow\ldots\rangle$, and these remain approximately degenerate for large but finite $\lambda$. This gives the ordered (ferromagnetic) phase. For $\lambda \equiv J_z/J_x\ll 1$, the transverse field dominates and energy is minimized when all spins are in the $+1$ eigenstate of $X$. There is a single ground state, $|\ldots ++ \ldots\rangle$, giving the disordered (paramagnetic) phase.\footnote{Here, and throughout, we will use arrows ($\uparrow,\downarrow$) to denote states in the $z$-basis and pluses/minuses ($+,-$) to denote states in the $x$-basis.}

The Hamiltonian given in (\ref{eq:ham_TFIM_open}) describes an open chain with free boundary conditions. Other boundary conditions are possible, such as periodic boundary conditions realized by adding an additional nearest neighbor  coupling for spin $N$ and spin $1$, as well as a different boundary condition compatible with duality which will be particularly useful to us. We will return to boundary conditions in a moment; for now we turn to the Kramers-Wannier duality of the TFIM.

\subsection{Kramers-Wannier duality}
Domain walls, here denoted by calligraphic operators and variables, offer a dual description of the one-dimensional TFIM. Living in between spins on a ``dual lattice'' labeled by half integer sites, they describe the frustration between spins on the original lattice. The existence of a domain wall at site $n+1/2$ is denoted by $\mathcal{X}$, defined as
\begin{equation} \label{eq:dual_tauX}
    \mathcal{X}_{n+1/2} = Z_n Z_{n+1}.
\end{equation}
If the spins on either side of the dual lattice site are aligned, $\mathcal{X}$ evaluates to $+1$, indicating the domain wall is turned off. A domain wall is on when the surrounding spins are anti-aligned, indicated by $\mathcal{X} = -1$.

A $\mathcal{Z}$ operator describes the creation (and annihilation) of domain walls. Flipping a single spin by applying the Pauli $X$ operator creates (or annihilates) two domain walls, one on either side of the spin. A single domain wall may be created by flipping all spins to one side of the dual lattice site, which we conventionally choose to be the right:
\begin{equation} \label{eq:dual_tauZ}
    \mathcal{Z}_{n-1/2} = \prod_{n'\geq n} X_{n'}.
\end{equation}
Because the product runs all the way to the end of the chain, this is a highly non-local operator in terms of the original variables. We note from these definitions that $\mathcal{X}$ and $\mathcal{Z}$ operators at the same dual lattice site $n+1/2$ anticommute:
\begin{equation}
    \big\{ \mathcal{X}_{n+1/2}, \, \mathcal{Z}_{n+1/2} \big\} = 0\,, \label{eq:tau_anticom}
\end{equation}
while $\mathcal{X}$ and $\mathcal{Z}$ operators at different dual lattice sites commute. Thus they obey the same Pauli matrix algebra as $X$ and $Z$, and the domain walls living at the dual lattice sites are also spin-$1/2$ variables.

These definitions may be inverted to find expressions for the Pauli $X$ and $Z$ operators in terms of the domain wall Pauli $\mathcal{X}$ and $\mathcal{Z}$ operators. Pauli $X$ may be written as a product of two $\mathcal{Z}$ operators,
\begin{equation} \label{eq:dual_X}
    X_n = \mathcal{Z}_{n-1/2} \mathcal{Z}_{n+1/2}.
\end{equation}
The domain wall equivalent of the Pauli $Z$ operator is defined analogously to (\ref{eq:dual_tauZ}) but with the product of all $\mathcal{X}$ operators on the opposite side of the lattice site, in our conventions the left:
\begin{equation} \label{eq:dual_Z}
    Z_n = \prod_{n' \leq n} \mathcal{X}_{n'-1/2}.
\end{equation}
Equations (\ref{eq:dual_tauX}), (\ref{eq:dual_tauZ}), (\ref{eq:dual_X}), and (\ref{eq:dual_Z}) describe the Kramers-Wannier duality transformation between spins and dual spins representing domain walls.

We will consider the effect of boundary conditions on the Kramers-Wannier duality shortly, but for the moment we obtain some intuition for the dual Hamiltonian by applying the transformations far from the boundary of the chain. Here, the TFIM Hamiltonian
\begin{equation}
    H_\text{TFIM} =  - J_z \big( \ldots + Z_{i-1} Z_i + Z_i Z_{i+1} + \ldots \big) - J_x \big( \ldots + X_i + X_{i+1} + \ldots \big)\,,
\end{equation}
is transformed to the domain wall Hamiltonian,
\begin{equation}
    H_\text{DW} = - \mathcal{J}_x \big( \ldots + \mathcal{X}_{i-1/2}+ \mathcal{X}_{i+1/2} + \ldots \big) - \mathcal{J}_z \big( \ldots + \mathcal{Z}_{i-1/2} \mathcal{Z}_{i+1/2} + \mathcal{Z}_{i+1/2} \mathcal{Z}_{i+3/2} + \ldots \big) ,
\end{equation}
where we have defined new couplings $\mathcal{J}_x \equiv J_z$ and $\mathcal{J}_z \equiv J_x$. The Hamiltonians are \textit{self-dual}, sharing the same form: two $\mathcal{Z}$ operators enforcing nearest neighbor coupling and single site $\mathcal{X}$ operators acting as a transverse field. The swapping of the couplings $J_x$ and $J_z$ implies that the dual coupling $\tilde\lambda \equiv \mathcal{J}_z/\mathcal{J}_x$ obeys $\tilde\lambda = 1/\lambda$, which is why the Kramers-Wannier duality is known as a ``strong-weak coupling duality''; in the regime where the original spins have nearest-neighbor couplings dominating over the transverse field $\lambda \equiv J_z/J_x\gg1$ their domain walls have the opposite, and vice versa. Acting the duality twice brings us back to the original variables and Hamiltonian, modulo boundary conditions to be discussed momentarily, and thus the original spins act as domain walls for their domain walls.

However, the full story must be slightly more subtle. We have already seen that for $\lambda \to \infty$ the TFIM has two-fold degenerate ground states, while for $\lambda \to 0$ there is a unique ground state. Thus it cannot be precisely strong-weak coupling self-dual, since the counting of ground states does not match. To see how this works precisely we have to think about boundary conditions. In the following subsections, we will consider three types of boundary conditions: open, periodic, and self-dual. The first two are reviewed primarily to gain intuition for the duality and to motivate the usefulness of self-dual boundary conditions, which will be implemented in the remainder of the paper. The reader may skim the review of open and periodic boundary conditions without missing any essential elements of the current work.

\subsubsection{Open boundary conditions}
\label{sec:open}
\begin{figure}
    \centering
    \includegraphics[width=0.7\linewidth]{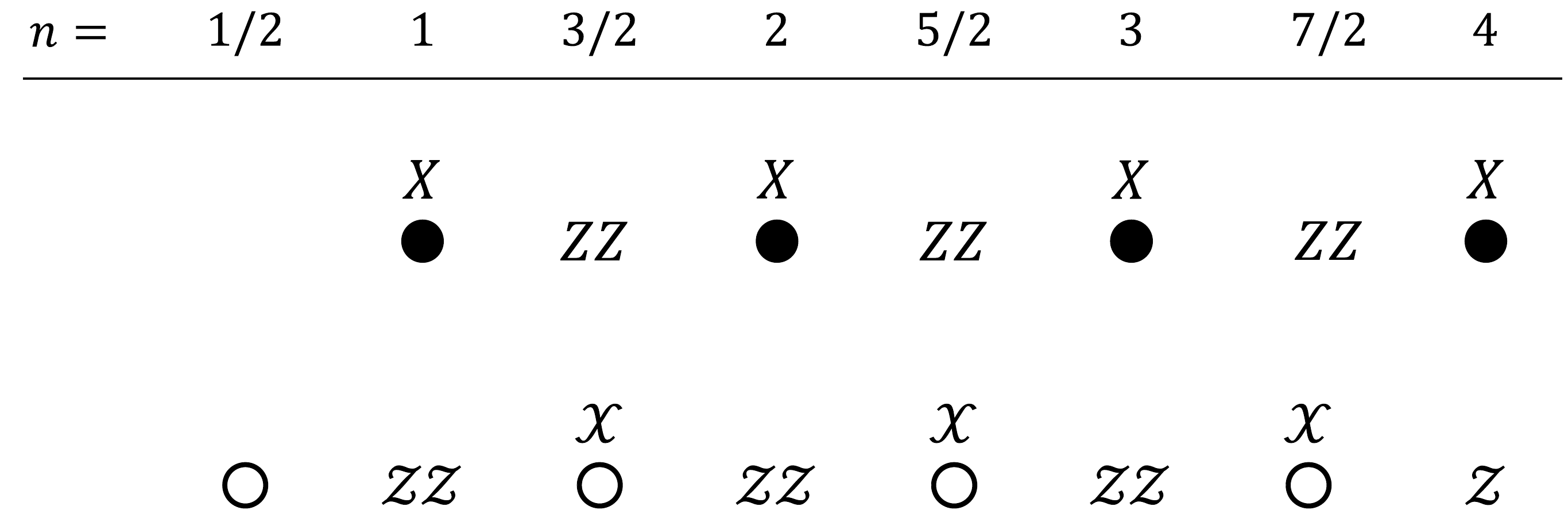}
    \caption{\centering{A pictorial representation of the spin Hamiltonian (\ref{eq:ham_TFIM_open}) and the dual domain wall Hamiltonian (\ref{eq:ham_DW_open}) for $N=4$. Solid dots indicate original lattice sites, while open dots indicate dual lattice sites.}}
    \label{fig:Spin_DW_duality}
\end{figure}

To see how the full Hilbert space acts under the Kramers-Wannier duality transformation, we must take care in defining the duality transformations on the boundaries of the chain. Consider the TFIM Hamiltonian (\ref{eq:ham_TFIM_open}) describing an open chain with free boundary conditions. The Hamiltonian for $N=4$,
\begin{equation}
    H_\text{TFIM}(N=4) =  - J_z \big( Z_1 Z_2 + Z_2 Z_3 + Z_3 Z_4 \big) -J_x \big(X_1 + X_2 + X_3 + X_4 \big)\,,
\end{equation}
is represented pictorially as the top row in Fig.~\ref{fig:Spin_DW_duality}. Each lattice site (represented by a solid dot) receives a Pauli $X$ operator, and a pair of two Pauli $Z$ operators connects each lattice site.

We now perform the duality transformation. Naturally, there are three dual lattice sites separating the four original lattice sites. This gives dual operators at $n=3/2,\,5/2,$ and $7/2$:
\begin{equation}
    \mathcal{X}_{3/2} = Z_1 Z_2, \quad \mathcal{X}_{5/2} = Z_2 Z_3, \quad \mathcal{X}_{7/2} = Z_3 Z_4 \label{eq:N=4_tauX_bulk}
\end{equation}
\begin{equation}
    \mathcal{Z}_{3/2} = X_2 X_3 X_4, \quad \mathcal{Z}_{5/2} = X_3 X_4, \quad \mathcal{Z}_{7/2} = X_4. \label{eq:N=4_tauZ_bulk}
\end{equation}
However, three spin-$1/2$ domain walls are not enough to describe the four spin-$1/2$ degrees of freedom in a state of the original spin chain; we must be missing a fourth dual lattice site. Let us try adding a site either at the beginning of the chain ($n=1/2$) or the end of the chain ($n=9/2$). Using (\ref{eq:dual_tauX}) and (\ref{eq:dual_tauZ}), the operators at the first site $n=1/2$ take the form
\begin{equation}
  \mathcal{X}_{1/2} = Z_1, \quad \mathcal{Z}_{1/2} = X_1 X_2 X_3 X_4 \label{eq:N=4_bdy} \,,
\end{equation}
where there is no $Z_0$ in the original chain so we took it to be equal to $\mathbb{1}$. These operators anticommute with each other and commute with the rest of the $\mathcal{X}_i$, $\mathcal{Z}_i$, so they successfully define a fourth spin-$1/2$ degree of freedom. If we had tried to define operators at the other end $n=9/2$ however, we would obtain $\mathcal{X}_{9/2} = Z_4$, $\mathcal{Z}_{9/2} = \mathbb{1}$, which commute and do not define a spin degree of freedom; so this is not a useful choice. Hence the correct procedure is to add a fourth domain wall site in front of the chain, at $n=1/2$, as in (\ref{eq:N=4_bdy}). Which end we added the lattice site was determined by our choice of direction convention in (\ref{eq:dual_tauZ}).

What is the meaning of this final dual degree of freedom? Unlike the others, it does not represent a domain wall for the original spins. Instead, since we have  $\mathcal{X}_{1/2} = Z_1$, it simply measures the value of the original degree of freedom at the left of the chain, and when this degree of freedom was $|\!\uparrow\rangle$ the dual $n=1/2$ degree of freedom is $|+\rangle$, while if the original degree of freedom was $|\!\downarrow\rangle$ the dual $n=1/2$ degree of freedom is $|-\rangle$. Thus the duality map on the full Hilbert space takes the set of $N$ spins to their $N-1$ domain walls, plus an extra degree of freedom simply recording the value of one of the spins on the boundary. Since the domain walls only measure relative orientation of the original spins, it is up to this final ``parity qubit" to measure the absolute orientation.

The full set of dual operators is given by equations (\ref{eq:N=4_tauX_bulk}), (\ref{eq:N=4_tauZ_bulk}), and (\ref{eq:N=4_bdy}). In terms of these dual operators, the Pauli $X$ and $Z$ operators for $N=4$ are given as
\begin{equation}
    X_1 = \mathcal{Z}_{1/2} \mathcal{Z}_{3/2}, \quad X_2 = \mathcal{Z}_{3/2} \mathcal{Z}_{5/2}, \quad X_3 = \mathcal{Z}_{5/2} \mathcal{Z}_{7/2}, \quad X_4 = \mathcal{Z}_{7/2} \label{eq:N=4_X}
\end{equation}
\begin{equation}
    Z_1 = \mathcal{X}_{1/2}, \quad Z_2 = \mathcal{X}_{1/2} \mathcal{X}_{3/2}, \quad Z_3 = \mathcal{X}_{1/2} \mathcal{X}_{3/2} \mathcal{X}_{5/2}, \quad Z_4 = \mathcal{X}_{1/2} \mathcal{X}_{3/2} \mathcal{X}_{5/2} \mathcal{X}_{7/2}. \label{eq:N=4_Z}
\end{equation} 
Finally, we obtain the domain wall Hamiltonian for $N=4$,
\begin{equation}
    H_\text{DW} (N=4) = - \mathcal{J}_x \big( \mathcal{X}_{3/2} + \mathcal{X}_{5/2} + \mathcal{X}_{7/2} \big) - \mathcal{J}_z \big( \mathcal{Z}_{1/2} \mathcal{Z}_{3/2} + \mathcal{Z}_{3/2} \mathcal{Z}_{5/2} + \mathcal{Z}_{5/2} \mathcal{Z}_{7/2} + \mathcal{Z}_{7/2} \big),
\end{equation}
where again we have used the dual couplings $\mathcal{J}_x = J_z$ and $\mathcal{J}_z = J_x$. This dual Hamiltonian is represented pictorially as the second line in Fig.~\ref{fig:Spin_DW_duality}, with open dots representing the dual lattice sites. Generalizing to arbitrary $N$ we find
\begin{equation} \label{eq:ham_DW_open}
    H_\text{DW, open} = -  \mathcal{J}_x\sum_{n=1}^{N-1}  \mathcal{X}_{n+1/2} - \mathcal{J}_z\left(\sum_{n=1}^{N-1} \mathcal{Z}_{n-1/2} \mathcal{Z}_{n+1/2} + \mathcal{Z}_{N-1/2} \right)\,.
\end{equation}
We note that the boundary terms for the dual Hamiltonian are different from the original Hamiltonian, in two ways. First, there is a single $\mathcal{Z}$ operator acting on the spin at the far right of the dual chain; this interaction, dual to the $X_N$ transverse field term, can be thought of as a nearest-neighbor coupling to a phantom spin off the edge of the dual chain which is always spin up. And second, the left-most dual spin lacks a transverse field coupling $\mathcal{X}$, since there was no nearest-neighbor coupling off the left edge of the chain in the original variables. Thus, the spin chain with open boundary conditions is only  ``self-dual up to boundary terms.''

This failure of exact self-duality is what is required to match the energy spectra of the two theories. The TFIM ground states were already enumerated below (\ref{eq:ham_TFIM_open}),
\begin{equation} \label{eq:TFIM_GDstates}
    \text{TFIM ground states} = 
        \begin{cases}
            |\uparrow\uparrow \dots \uparrow\rangle_z, \, |\downarrow\downarrow \dots \downarrow\rangle_z \quad &\lambda \to \infty\\
            |++ \dots + \rangle_x & \lambda \to 0 \,.
        \end{cases}
\end{equation}
In the $\lambda \equiv J_z/J_x \gg 1$ limit, the first excited states of the TFIM are given by frustrating only one pair of nearest neighbors, such as $|\!\uparrow \downarrow \downarrow \dots \downarrow\rangle_z$ or $|\!\uparrow \uparrow \downarrow \dots \downarrow \rangle_z$. In the opposite limit $\lambda \gg 1$, first excited states are given by flipping single sites only: $|-++\dots +\rangle_x$ or $|+-+\dots +\rangle_x$ for example. The highest energy TFIM states are given by maximum frustration when $J_z$ dominates, say $|\uparrow\downarrow\uparrow\downarrow\dots\uparrow\downarrow\rangle_z$. When $J_x$ dominates, the highest energy state can be found by placing all spins in opposition to the transverse field, $|---\dots-\rangle_x$.

Now consider the  energy spectrum for the domain wall Hamiltonian (\ref{eq:ham_DW_open}), beginning with the ground states. In the $\tilde\lambda \equiv  \mathcal{J}_z / \mathcal{J}_x \ll 1$ limit, energy is minimized when all domain walls at dual sites $n \geq 3/2$ are in the $+1$ eigenvalue of $\mathcal{X}$. Since $\mathcal{X}_{1/2}$ does not appear in the Hamiltonian (\ref{eq:ham_DW_open}), the first site does not contribute to the energy of the state, and so both orientations of this spin lead to ground states. Therefore there are two ground states in this limit. In the opposite limit, $\tilde\lambda \equiv \mathcal{J}_z / \mathcal{J}_x \gg 1$, nearest neighbor sites want to be mutually aligned in the $\mathcal{Z}$ direction. Due to the presence of the single $\mathcal{Z}$ in (\ref{eq:ham_DW_open}), the $+1$ eigenstate of $\mathcal{Z}$ is preferred, aligning all sites in this direction. This leads to a single ground state in this limit. The ground states are then
\begin{equation} \label{eq:DW_GDstates}
    \text{DW ground states} = 
        \begin{cases}
            |++\dots+\rangle_{\mathcal{X}}, \, |-+\dots+\rangle_{\mathcal{X}} \quad & \tilde\lambda \to 0 \\
            |\uparrow\uparrow \dots \uparrow\rangle_{\mathcal{Z}} \quad & \tilde\lambda \to \infty.
        \end{cases}
\end{equation}
The first excited states are similar to those of the TFIM, given by turning one domain wall on when $\mathcal{J}_x$ dominates (e.g.\ $|\pm + - + \dots + \rangle_{\mathcal{X}}$) and creating one frustrated pair when $\mathcal{J}_z$ dominates (e.g.\ $|\uparrow\uparrow\downarrow\dots\downarrow\rangle_{\mathcal{Z}}$). The highest energy states are also similar: $|\pm-\dots-\rangle_{\mathcal{X}}$ for $\mathcal{J}_x / \mathcal{J}_z \gg 1$ and $|\uparrow\downarrow\uparrow\downarrow\dots\uparrow\downarrow\rangle_{\mathcal{Z}}$ when $\mathcal{J}_x / \mathcal{J}_z \ll 1$, where for the latter case the domain wall located at $n=N-1/2$ must be in the $-1$ eigenstate of the $\mathcal{Z}$ operator to be antialigned with the phantom spin off the end of the chain. The TFIM and DW theories have the same number of states in the same limits of $J_z / J_x = \mathcal{J}_x/\mathcal{J}_z$ at all levels of the energy spectrum. This should be expected, as the duality only relabels the degrees of freedom, without changing the underlying physics.

Thus the different boundary terms in the domain wall Hamiltonian (\ref{eq:ham_DW_open}) can be understood as reconciling the energy spectra between the two theories. The absence of a single ${\mathcal X}$ term introduces a two-fold degeneracy at large dual coupling, while the single $\mathcal{Z}$ term  lifts the degeneracy at small dual coupling, precisely what is needed to map the spectrum of the dual theory to the original.

\subsubsection{Periodic boundary conditions}
\label{sec:periodic}

Other boundary conditions are possible. One natural choice is to effectively place the spins of the TFIM on a ring or circle by introducing a nearest neighbor coupling between the $1^\text{st}$ and $N^\text{th}$ spins. For general $N$, the periodic Hamiltonian is given by
\begin{equation} \label{eq:ham_TFIM_per}
    H_\text{TFIM, per.} =-J_z\left(\sum_{n=1}^{N-1} Z_n Z_{n+1}+ Z_NZ_1\right)- J_x\sum_{n=1}^N  X_n \,.
\end{equation}
The new coupling does not change the ground states at $\lambda \to \infty$, but it does modify the excited states; a configuration like $|\! \! \uparrow \cdots \uparrow \downarrow \cdots \downarrow \rangle$ now counts as two domain walls, one in the middle and one between the first and last sites in the chain, and thus receives twice the energy above the ground state.

We must now choose how to implement KW duality for this case. If we truly think of the spins as living on a circle, the definition (\ref{eq:dual_tauZ}) for $\mathcal{Z}$ might seem to loop around infinitely on itself and be ill-defined. Instead, we can opt to use the same definitions of the $\mathcal{Z}$ as with open boundary conditions, cutting off the product after site $N$; despite the circle symmetry of the periodic Hamiltonian, we choose a special link on the chain to define duality. Thus compared to the open boundary conditions we are only changing the Hamiltonian, and leave the duality map on the state space alone.

Using the same dual operator definitions, the new $Z_N Z_1$ term in (\ref{eq:ham_TFIM_per}) leads to a non-local term in the domain wall Hamiltonian:
\begin{equation} \label{eq:ham_DW_per}
    H_\text{DW, per.} =  -  \mathcal{J}_x\left(\sum_{n=1}^{N-1}  \mathcal{X}_{n+1/2}+  \prod_{n\geq 1} \mathcal{X}_{n+1/2}\right) - \mathcal{J}_z\left(\sum_{n=1}^{N-1} \mathcal{Z}_{n-1/2} \mathcal{Z}_{n+1/2} + \mathcal{Z}_{N-1/2} \right)\,,
\end{equation}
where again $\mathcal{J}_z = J_x$ and $\mathcal{J}_x = J_z$. The non-locality of the new term is another consequence of choosing a special point to represent the beginning of the chain. It provides an essential correction to the domain wall energy spectrum: configurations with an odd number of domain walls turned on receive an extra energy penalty. This accounts for the domain wall between the $1^\text{st}$ and $N^\text{th}$ spins, which is effectively excluded from the state space by our choice of a special link in the chain. Thus duality again preserves the energy spectrum, as it must.

Note in this discussion we have not imposed any periodicity constraint on the Hilbert space, only on the Hamiltonian; since spins can flip between any two sites, it is natural not to impose any matching constraint between the first and last sites in the space of states either. One could imagine other choices; for example, \citep{radicevic_spin_2019} enforces periodicity in the state space rather than the Hamiltonian. In doing so, no site is chosen as the ``start'' of the chain, and the ring is left unbroken; this leads to self-dual periodic boundary conditions, at the cost of requiring a projection on the Hilbert space for the duality to match. We will instead opt to achieve self-dual boundary conditions on an open chain, which does not require any projection on the space of states, which we turn to now.

\subsubsection{Self-dual boundary conditions} \label{sec:selfdualBCs}
None of the boundary conditions presented so far have realized self-dual Hamiltonians; in each case, self-duality is contaminated by boundary terms. While these boundary terms should disappear in the thermodynamic limit, one might wish to consider self-duality in a finite chain. Here we present a modification of the open chain boundary conditions that remedies this.

The dual to the open chain with free boundary conditions possessed two modifications: a missing transverse field on one end site to add degeneracy at strong dual coupling, and a phantom nearest neighbor interaction on the other end site to remove degeneracy at weak dual coupling. In fact if we consider a Hamiltonian with only one of these modifications, it will be fully  self-dual. Depending on which we choose, we can have a theory with ground state degeneracy at all couplings, or one with a unique ground state at all couplings. Let us focus on the former.

Consider a TFIM Hamiltonian with open boundary conditions, and remove the transverse field $X_N$ at site $N$. This Hamiltonian is
\begin{equation}\label{eq:Hamiltonian_self_dual}
    H_\text{TFIM, self-dual} = -J_z   \sum_{n=1}^{N-1} Z_n Z_{n+1} - J_x \sum_{n=1}^{N-1} X_n \,.
\end{equation}
Acting with duality, the removal of the $X_N$ term removes the single dual $\mathcal{Z}_{N-1/2}$ from the dual Hamiltonian. The $n=1/2$ site in the dual lattice still has no $\mathcal{X}_{1/2}$ term since there is still no nearest-neighbor interaction off the left edge of the original chain. We obtain
\begin{equation}
    H_\text{DW, self-dual} =  -  \mathcal{J}_x\sum_{n=1}^{N-1}  \mathcal{X}_{n+1/2} - \mathcal{J}_z \sum_{n=1}^{N-1} \mathcal{Z}_{n-1/2} \mathcal{Z}_{n+1/2}\,.
\end{equation}
The Hamiltonians have the same form, up to which end has the site without a transverse field operator. Thus if we define the full duality to be composition of Kramers-Wannier duality and a parity flip along the chain, the theory is then precisely self-dual.

The self-dual Hamiltonian has an exact SU(2) symmetry preserved at all values of the coupling $\lambda$, generated by the operators 
\begin{equation}
    X_1 X_2 \ldots X_N, \quad X_1 X_2 \ldots X_{N-1} Y_N, \quad Z_N \,,
\end{equation}
which commute with $ H_\text{TFIM, self-dual}$. Due to the presence of the symmetry, the entire energy spectrum is two-fold degenerate at all values of $\lambda$. In particular, the degenerate  ground states are given by
\begin{equation}\label{eq:SelfDualGdState}
    \text{Self-dual TFIM ground states} = 
        \begin{cases}
            |\!\uparrow\uparrow \dots \uparrow\rangle_z, \, |\!\downarrow\downarrow \dots \downarrow\rangle_z \quad &\lambda \to \infty \\
            |++\dots++\rangle_x, \, |++\dots+-\rangle_x \quad &\lambda \to 0
        \end{cases}
\end{equation}
The  ground states at strong and weak coupling are eigenvectors of the $SU(2)$ generators  $Z_N$ or $X_1\ldots X_N$, respectively, and those at  intermediate coupling are eigenvectors of some linear combination of these generators.

We note for completeness the complementary self-dual boundary conditions, adding only the coupling to a phantom nearest neighbor on one end of the chain,
\begin{equation}
H_\text{DW, self-dual'} = - J_z \left( \sum_{n=1}^{N-1} Z_n Z_{n+1}+ Z_N \right) -J_x \sum_{n=1}^N X_n\,,
\end{equation}
which is also exactly self-dual, and due to the lack of $SU(2)$ symmetry possesses a unique ground state for all values of the coupling.

To summarize, we have seen how Kramers-Wannier duality maps a Hilbert space of $N$ spins to a Hilbert space of $N$ dual spins, which correspond to the $N-1$ domain walls of the original variables, plus a final ``parity qubit" degree of freedom that just corresponds to one of the original spins on the end of the chain. This map on the space of states is independent of the precise form of the Hamiltonian. Transverse field Ising model Hamiltonians with open or periodic boundary conditions are self-dual under this transformation only up to boundary terms, but we demonstrated the existence of transverse field Ising Hamiltonians with exactly self-dual boundary conditions as well.

\subsection{Local entanglement under Kramers-Wannier duality} \label{sec:local_EE}

We now turn to the study of entanglement entropy and its transformation under duality. As usual, this entails choosing a state $\rho$ in the system, and choosing some subset $B$ of the degrees of freedom of the theory, and tracing them out to produce a density matrix for the remaining degrees of freedom $A$,
\begin{equation}
    \rho_A = \text{Tr}_B\, \rho \,,
\end{equation}
and then calculating the von Neumann entropy for $\rho_A$\,,
\begin{equation}
    S(A) = - \text{Tr}_A \, \rho_A \log \rho_A \,. 
\end{equation}
If the bipartite state $\rho$ is pure, then $S(A) = S(B)$. A common choice of state to calculate the entanglement entropy for is the ground state. A simple case is to let $A$ and $B$ be two connected blocks of sites, $A$ to the left and $B$ to the right. In the limit of an infinite chain $N \to \infty$ in both directions, which site is chosen to be the boundary between the two regions becomes irrelevant as all choices are equivalent.

Let us start by considering open boundary conditions, but our results will hold true for all boundary conditions. As we have described, for the TFIM with open boundary conditions (\ref{eq:ham_TFIM_open}) for any finite $\lambda$ there is a unique ground state; at $\lambda = 0$ we know this is just  
\begin{equation} \label{eq:SelfDualEntanglement}
    |\psi_0\rangle_\text{TFIM}(\lambda = 0) =   |++\dots+\rangle_x \,.      
\end{equation}
This is an unentangled state, and tracing out a block of spins $B$ will leave a pure state with zero entanglement entropy.

\begin{figure}
    \centering
    \includegraphics[width=0.7\linewidth]{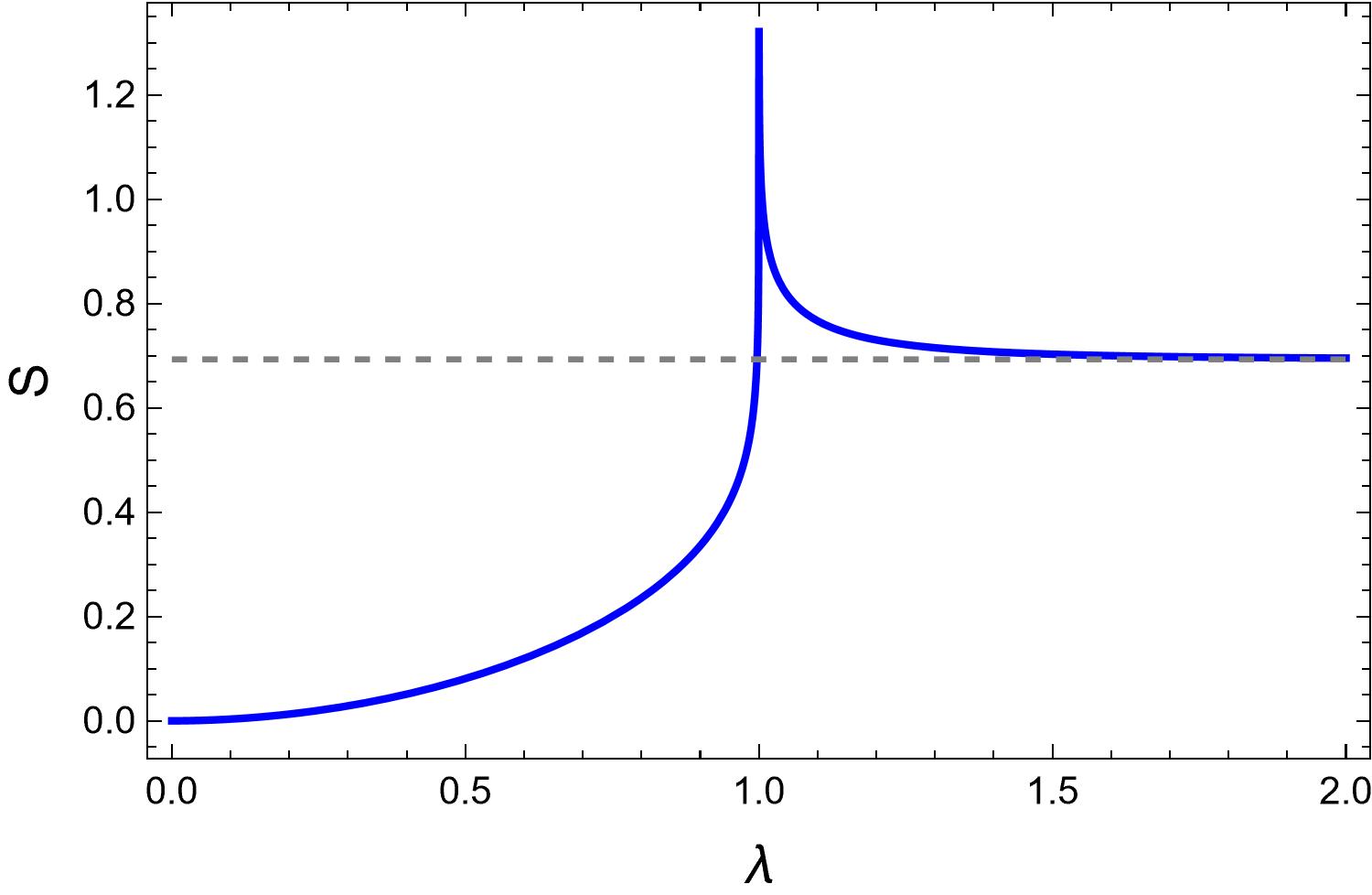}
    \caption{\centering{A plot of the entanglement entropy $S$ for two halves of the 1D TFIM calculated by \citep{Calabrese:2004eu} as a function of the dimensionless ratio $\lambda = J_z / J_x$ (solid blue curve). A gray dashed line marks $S=\ln 2$.}}
    \label{fig:TFIM_EE_Calabrese}
\end{figure}

On the other hand, for $J_x = 0$ the open boundary conditions spectrum becomes degenerate, but one can determine the appropriate linear combination of the degenerate states that is approached in the $\lambda \to \infty$ limit using degenerate perturbation theory. Treating the single site $X$ operators as a perturbation on the nearest neighbor $ZZ$ couplings and diagonalizing this perturbation in the degenerate subspace spanned by  $|\!\uparrow\uparrow\dots\uparrow\rangle_z$ and  $|\!\downarrow\downarrow\dots\downarrow\rangle_z$ for a chain of $N$ spins lifts the degeneracy at $N^\text{th}$ order, since all $N$ Pauli $X$ operators are needed to flip the $N$ spins from the $\pm1$ to $\mp1$ eigenstates of $Z$. One finds the linear combination
\begin{equation} \label{eq:TFIM_entangled}
    |\psi_0\rangle_\text{TFIM}(\lambda \to \infty) = \frac{1}{\sqrt{2}} \big( |\!\uparrow\uparrow\dots\uparrow\rangle_z + |\!\downarrow\downarrow\dots\downarrow\rangle_z \big)\,,           
\end{equation}
is the ground state approached in the $J_z / J_x \gg 1$ limit. This state is manifestly entangled for any $N \geq 2$. Tracing out a block of spins $B$ leaves a mixed density matrix
\begin{equation} \label{eq:finite_reduced}
    \rho_A(\lambda \to \infty) = \frac{1}{2} \left(|\!\uparrow\uparrow\dots\uparrow\rangle_z \langle\uparrow\uparrow\dots\uparrow\!\!|  + |\!\downarrow\downarrow\dots\downarrow\rangle_z \langle\downarrow\downarrow\dots\downarrow\!\!| \right) 
\end{equation}
leading to entanglement entropy $S(A) = \ln 2$, with the value 2 coming from the two maximally entangled states of (\ref{eq:TFIM_entangled}). The entanglement entropy for the ground state of the TFIM with open boundary conditions was calculated in the thermodynamic ($N \to \infty$) limit  for all $\lambda$  by \citep{Calabrese:2004eu} and the result is shown in Fig.~\ref{fig:TFIM_EE_Calabrese}, where the limits $S(\lambda \to0) = 0 $ and $S(\lambda \to \infty) = \ln 2$ are both visible. 
For generic values of $\lambda$ the ground state is entangled, and there is a cusp at the value $\lambda = 1$.

We now turn to the transformation of local entanglement entropy in the TFIM under Kramers-Wannier duality.  We consider the ground states of the domain wall Hamiltonian (\ref{eq:ham_DW_open}). There is a unique ground state for any nonzero $\mathcal{J}_z$, and in the limit $\tilde\lambda \to \infty$ the ground state is
\begin{equation} 
    |\psi_0\rangle_\text{DW}(\tilde\lambda \to \infty) =   |\!\!\uparrow\uparrow\dots\uparrow\rangle_{\mathcal{Z}} \,.           
\end{equation}
This is dual to the state $|++\dots+\rangle_x$, and like it is unentangled; thus tracing out any block of spins will give zero entanglement entropy. Thus for this pair of dual states local entanglement is preserved.

However, in the limit $\tilde\lambda\to 0$ the story is different. There the spectrum is degenerate, but we can again use degenerate perturbation theory to find the combination that is approached in the limit; alternately we can apply the duality map to the state (\ref{eq:TFIM_entangled}). Either way we obtain
\begin{eqnarray}
    |\psi_0 \rangle_\text{DW}(\tilde\lambda \to0) &=& \frac{1}{\sqrt{2}} \big( |++\dots+\rangle_\mathcal{X} + |-+\dots+\rangle_\mathcal{X} \big) \nonumber \\
        &=& |\uparrow\rangle_\mathcal{Z} |+\dots+\rangle_\mathcal{X}, \quad J_z / J_x \gg 1 \label{eq:DW_unentangled}
\end{eqnarray}
which  factors into a product state; this domain wall ground state is unentangled and under tracing out any block of dual spins has zero entanglement entropy. Thus we see the entangled TFIM ground state (\ref{eq:TFIM_entangled}) is related to the unentangled domain wall ground state (\ref{eq:DW_unentangled}) by Kramers-Wannier duality, and hence local entanglement is not preserved by duality.

The issue is that while the TFIM states $ |\!\uparrow\uparrow\dots\uparrow\rangle_z$ and $|\!\downarrow\downarrow\dots\downarrow\rangle_z$ differ in every spin in the original degrees of freedom, their domain wall structure is identical,  all domain walls being off in both cases, with only the overall orientation being opposite. In the dual (domain wall) variables, this means all degrees of freedom but one   (the ``parity qubit") are identical, and the dual states $|++\dots+\rangle_\mathcal{X}$ and $|-+\dots+\rangle_\mathcal{X}$ differ only in the first dual spin, and thus linear combinations of them are unentangled. This is possible because the duality map is non-local.

This result is not unique to using open boundary conditions. In fact for all our boundary conditions we used the same duality map, and since entanglement is a statement about the states and not their energies, the same result holds for these states with periodic or self-dual boundary conditions as well. The state (\ref{eq:DW_unentangled}) requires an additional flip of the order of the spins be included in the action of duality to match the self-dual boundary condition ground states (\ref{eq:SelfDualGdState}), so in this case we compare the dual ground states
\begin{equation}\label{eq:Self_dual_ground_states}
  |\psi_0\rangle_{\lambda \to 0 }=   |+\cdots + \uparrow\rangle  \quad \longleftrightarrow \quad |\psi_0\rangle_{\lambda \to \infty } = \frac{1}{\sqrt{2}} \big( |\!\uparrow\uparrow\dots\uparrow\rangle + |\!\downarrow\downarrow\dots\downarrow\rangle \big) \,,
\end{equation}
where again local entanglement is present on one side of the duality, and absent on the other side.

It is natural to wonder what has happened to the entanglement after the duality. Is it gone? The issue is that we are using a new set of degrees of freedom, and it is these other degrees of freedom that are not entangled. If the entanglement is still ``there", it is  present in other, non-local degrees of freedom correseponding to the original spins. The goal of the rest of the paper is to explore ways to characterize this non-local entanglement.

\section{Non-local degrees of freedom from quantum channels}
\label{sec:QChannels}

We ordinarily characterize degrees of freedom in our Hilbert space by factorizing it as ${\cal H} = {\cal H}_1 \otimes{\cal H}_2 \otimes \cdots$, with each ${\cal H}_i$ corresponding to a degree of freedom. Each degree of freedom may be thought of as ``near" certain other degrees of freedom; in a spin model like the transverse field Ising model this locality is ultimately determined by the Hamiltonian in the form of the nearest neighbor couplings, but in a continuum theory we typically associate the degrees of freedom with points in a space whose topology determines locality. As we have seen, however, duality is a non-local transformation that rearranges the degrees of freedom, and entanglement that is manifest can become hidden in non-local degrees of freedom. We would like to have an explicit way to characterize degrees of freedom of our quantum systems, even if the degrees of freedom are not the usual ``local" ones. 

The intuitive idea of a single degree of freedom is a way in which the state of the system may change, while all other aspects of the state stay the same. We would like to have an operator or set of operators associated with a single degree of freedom in the system changing; we can begin by characterizing ordinary, local degrees of freedom, but in a way we can generalize to the non-local case. It turns out a useful way to do this is in the language of quantum channels.

A useful starting point is the notion of the partial trace.  Calculating the entanglement entropy of the $A$ subsystem requires that we ``forget" the $B$ degrees of freedom by performing a partial trace over the degrees of freedom contained in the $B$ subsystem,
\begin{eqnarray}
\label{eq:PartialTrace}
    \rho_A &=& \tr_{B} \rho_{AB} \nonumber \\
        &=& \sum_{b \in B} \big( \mathbb{1}_A \otimes \langle b |_B \big)\rho_{AB} \big( \mathbb{1}_A \otimes | b \rangle_B \big)\,,
\end{eqnarray}
where $\{|b\rangle\}$ are a set of orthonormal basis vectors for $B$. It is convenient for us to express the partial trace as a special case of a quantum operation, also called a quantum channel.

\subsection{Review of quantum channels}

A quantum channel, or quantum operation, $\mathcal{N}(\rho)$ is a linear map on density matrices that preserves hermiticity, trace,\footnote{It can be useful to consider non-trace preserving quantum channels, but we will not need to do so.} and positivity, even when $\rho$ is just one factor in a larger density matrix, and thus maps density matrices to density matrices. Any quantum channel $\mathcal{N}(\rho)$ acting on a density matrix $\rho$ can be expressed by the action of an isometry $V$ followed by a partial trace,
\begin{equation}
    \mathcal{N}(\rho) = \tr_B \big( V \rho V^\dagger \big)\,, \label{eq:gen_channel}
\end{equation}
where the isometry may be thought of as a combination of tensoring in new degrees of freedom to the Hilbert space, for example an environment that an open system is coupled to, and performing a unitary transformation on the enlarged space. Thus a quantum channel generalizes unitary evolution of a system to include adding and removing degrees of freedom.

Any quantum channel can be written in an \textit{operator-sum representation},
\begin{equation}
    \mathcal{N}(\rho) = \sum_b K_b \rho K_b^\dagger\,,
\end{equation}
where the {\em Kraus operators} $K_b$ are related to the isometry $V$ as
\begin{equation}
    V = \sum_b |b\rangle_B \otimes K_b\,,
\end{equation}
with $\{|b\rangle\}$  an orthonormal basis on $B$.
The isometry property $V^\dagger V = \mathbb{1}$ guarantees the Kraus operators obey 
\begin{equation}
\label{eq:KrausSum}
   \sum_b K_b^\dagger K_b=  \mathbb{1}\,.
\end{equation} 
This and the fact that the $K^\dagger_b K_b$ are Hermitian and positive indicates that they form a positive operator-valued measure (POVM) partitioning the Hilbert space.

The same quantum operation may be defined by more than one set of Kraus operators. If we change the basis $\{|b\rangle\}$ for $B$ by a unitary transformation $U$, the Kraus operators will be transformed as well, as
\begin{equation}\label{eq:Kraus_basis}
    K'_b = U_b^a K_a \,.
\end{equation}
Any two sets of Kraus operators $\{ K_b\}$, $\{ K_b'\}$ related in this way describe the same quantum channel. The composition of two quantum channels $\{ K_b\}$, $\{ L_j\}$  is also a quantum channel, with Kraus operators $\{ L_j K_b\}$.

In general the output density matrix $\mathcal{N}(\rho)$ may act on a Hilbert space that is larger, smaller, or the same size as the Hilbert space acted on by $\rho$, depending on how many degrees of freedom are added by the isometry $V$ and how many are removed by the trace. A common application is for $\mathcal{N}(\rho)$ and $\rho$ to act on the same Hilbert space, where the degrees of freedom tensored in by $V$ are the same degrees of freedom traced out at the end; this is the way to describe the evolution of an open system in contact with an environment. However, we will be interested in quantum channels that remove degrees of freedom, and thus reduce the size of the Hilbert space.

Let us specialize to the case where no degrees of freedom are tensored in, and thus the isometry $V$ becomes a unitary, and the whole quantum channel is just the action of this unitary followed by a partial trace. The Kraus operators for such a channel map the Hilbert space ${\cal H}_{AB}$ to ${\cal H}_A$. A unitary satisfies $VV^\dagger = \mathbb{1}$ as well as $V^\dagger V = \mathbb{1}$, and thus in addition to $ \sum_b K_b^\dagger K_b=  \mathbb{1}_{AB}$ (\ref{eq:KrausSum}) the Kraus operators obey
\begin{equation}
    K_{b'} K_b^\dagger = \mathbb{1}_A \delta_{bb'}. \label{eq:KK_cond}
\end{equation}
The condition (\ref{eq:KK_cond}) implies that each $K_b^\dagger$ is itself an isometry. It also indicates that the $K^\dagger_b K_b$ that comprise the POVM (\ref{eq:KrausSum})  obey
\begin{equation}
    (K^\dagger_b K_b)(K^\dagger_{b'}K_{b'}) = \delta_{bb'} K^\dagger_b K_b \,,
\end{equation}
and are thus orthogonal projectors.

\subsection{Partial trace and local degrees of freedom}
\label{sec:KrausMaps}

Let us now consider the partial trace of some degrees of freedom $B$ (\ref{eq:PartialTrace}) as a quantum channel. In this case the isometry is trivial, $V = \mathbb{1}_{AB}$. The Kraus operators mapping the Hilbert space from ${\cal H}_{AB}$ to ${\cal H}_A$ take the form
\begin{equation}
    K_b = \mathbb{1}_A \otimes \langle b |_B   \,.
\end{equation}
Each $K_b$ finds and removes the corresponding $B$ basis vector $|b\rangle$ while leaving $A$ alone, and hence the quantum channel generates a term in the reduced density matrix $\rho_A$ for each value $b$. If the original state is entangled between $A$ and $B$, this density matrix must have multiple terms and will thus be mixed.

The adjoint of $K_b$ is a map from ${\cal H}_A$ to ${\cal H}_{AB}$,
\begin{equation}
    K_b^\dagger = \mathbb{1}_A \otimes | b \rangle_B   \,,
\end{equation}
and we may think of it as a kind of creation operator, adding a new tensor factor $B$ to the Hilbert space in the state $|b \rangle_B$. This is consistent with it being an isometry as described below (\ref{eq:KK_cond}). The Kraus operator $K_b$ is then the corresponding annihilation operator, removing the Hilbert space factor ${\cal H}_B$ and leaving whatever state in ${\cal H}_A$ that was in a tensor product with $|b\rangle_B$.

From the set of partial trace Kraus operators we can construct the matrix of operators on ${\cal H}_{AB}$,
\begin{equation}\label{eq:Tmatrix}
    T_{b'b} \equiv  K^\dagger_{b'} K_b \,.
\end{equation}
The off-diagonal operators $T_{b'b}$ may be thought of as raising/lowering operators, replacing a state with $B=b$ with a state with $B=b'$, while leaving the $A$ degrees of freedom alone. This is in contrast to $K_b$ alone, which annihilates the ${\cal H}_B$ factor in the Hilbert space entirely. The diagonal operators $T_{bb}$ are, as mentioned, a set of projectors onto the subspace where $B=b$.

Thus we see that we may characterize the degrees of freedom $B$ by the Kraus operators $K_b$, which are in one-to-one correspondence with the possible values of $B$ in a particular basis. Moreover, they act to create/annihilate/modify/project on the value of $B$ while leaving the value of $A$ untouched, a property we want in characterizing $B$ as a degree of freedom independent of $A$.

It is useful to consider an example. Let $A$ and $B$ each be single qubits, so ${\cal H}_{AB}$ is a two-qubit system. In a basis $\{|\!\uparrow\uparrow\rangle,  |\!\downarrow\uparrow\rangle, |\!\uparrow\downarrow\rangle, |\!\downarrow\downarrow\rangle\}$, the Kraus operators are
\begin{eqnarray}
    K_\uparrow =& \mathbb{1}_A \otimes \langle \uparrow \!|_B =& 
        \begin{pmatrix}
            1   &   0   &   0   &   0 \\
            0   &   1  &   0   &   0
        \end{pmatrix} \label{eq:vup} \\
    K_\downarrow =& \mathbb{1}_A \otimes \langle \downarrow \!|_B =&
        \begin{pmatrix}
            0   &   0   &   1   &   0 \\
            0   &   0   &   0   &   1
        \end{pmatrix}. \label{eq:vdn}
\end{eqnarray}
We can see that $K_\uparrow$ picks out the first two entries in a column matrix, for which the second spin is up, while $K_\downarrow$ picks out the last two entries in a column matrix, for which the second spin is down.

The corresponding creation operators are
\begin{equation}
    K_\uparrow^\dagger = \mathbb{1}_A \otimes | \!\uparrow \rangle_B  = 
        \begin{pmatrix}
            1   &   0    \\
             0   &   1    \\
             0   &   0   \\
             0   &   0    
        \end{pmatrix} \,, \quad \quad
    K_\downarrow^\dagger = \mathbb{1}_A \otimes | \!\downarrow \rangle_B =
         \begin{pmatrix}
            0   &   0    \\
             0   &   0    \\
             1   &   0   \\
             0   &   1    
        \end{pmatrix} \,. \label{eq:kdaggerupdn}
\end{equation}
acting on a single spin (subsystem $A$) to give a two spin state:
\begin{eqnarray}
    K_\uparrow^\dagger |\!\uparrow\rangle_A = |\!\uparrow\uparrow\rangle,&& \quad K_\uparrow^\dagger |\!\downarrow\rangle_A = |\!\downarrow\uparrow\rangle \\
    K_\downarrow^\dagger |\!\uparrow\rangle_A = |\!\uparrow\downarrow\rangle,&& \quad K_\downarrow^\dagger |\!\downarrow\rangle_A = |\!\downarrow\downarrow\rangle.
\end{eqnarray}
The POVM forming the projectors onto the subspaces of definite eigenvalues for $B$ are
\begin{eqnarray}
  T_{\uparrow\uparrow} =&  K_\uparrow^\dagger K_\uparrow =& \mathbb{1}_A \otimes | \!\uparrow \rangle_B \langle \uparrow \!|_B  = 
        \begin{pmatrix}
            1   &   0   & 0 & 0 \\
             0   &   1 &0 &0   \\
             0   &   0 &0 &0  \\
             0   &   0  &0 &0  
        \end{pmatrix} \,, \label{eq:tUU} \\
   T_{\downarrow\downarrow} =&  K_\downarrow^\dagger K_\downarrow =& \mathbb{1}_A \otimes | \!\downarrow \rangle_B \langle \downarrow \!|_B =
         \begin{pmatrix}
          0 &0 & 0   &   0    \\
         0 &0 &   0   &   0    \\
          0 &0&   1   &   0   \\
          0 &0 &  0   &   1    
        \end{pmatrix} \,. \label{eq:tDD}
\end{eqnarray}
and the raising/lowering operators are
\begin{eqnarray}
  T_{\uparrow\downarrow} =&  K_\uparrow^\dagger K_\downarrow =& \mathbb{1}_A \otimes | \!\uparrow \rangle_B \langle \downarrow \!|_B  = 
        \begin{pmatrix}
            0   &   0   & 1 & 0 \\
             0   &   0 &0 &1   \\
             0   &   0 &0 &0  \\
             0   &   0  &0 &0  
        \end{pmatrix} \,, \label{eq:tUD} \\
   T_{\downarrow\uparrow} =&  K_\downarrow^\dagger K_\uparrow =& \mathbb{1}_A \otimes | \!\downarrow \rangle_B \langle \uparrow \!|_B =
         \begin{pmatrix}
          0 &0 & 0   &   0    \\
         0 &0 &   0   &   0    \\
          1 &0&   0   &   0   \\
          0 &1 &  0   &   0    
        \end{pmatrix} \,. \label{eq:tDU}
\end{eqnarray}
It is easy to see that these operators change or project onto a value of $B$ while leaving $A$ alone.

In general, if we write the column vectors for our Hilbert space ${\cal H}_{AB} = {\cal H}_A \otimes {\cal H}_B$ with dimensions $d_A$, $d_B$ as
\begin{equation}
    |\psi \rangle = \begin{pmatrix}
        a_1 b_1 \\ a_2 b_1 \\ \vdots  \\ a_1 b_2 \\ a_2 b_2\\ \vdots
    \end{pmatrix}\,,
\end{equation}
then the Kraus operators $K_i$ take the form
\begin{eqnarray} \nonumber
    K_1 &=& 
        \begin{pmatrix}
            \mathbb{1}_{d_A \times d_A} & 0_{d_A \times d_A} & \cdots & 0_{d_A \times d_A} 
        \end{pmatrix}    \\ 
    K_2 &=& 
        \begin{pmatrix}
            0_{d_A \times d_A} & \mathbb{1}_{d_A \times d_A} &  \cdots & 0_{d_A \times d_A} 
        \end{pmatrix}   \\ \nonumber
    &\vdots& \\
    K_{d_B} &= &
        \begin{pmatrix}
            0_{d_A \times d_A} & 0_{d_A \times d_A} &  \cdots &  \mathbb{1}_{d_A \times d_A}
        \end{pmatrix} \,, \nonumber
\end{eqnarray}
where there are a total of $d_B$ blocks, each of size $d_A \times d_A$, making up each Kraus operator. Again it is fully manifest that these operators act on the $B$ values but not the $A$ values.

In the above we imagined tracing out the second ($B$) degree of freedom. We can just as easily write down Kraus operators $K^A_a$ to trace out the first ($A$) degree of freedom,
\begin{eqnarray}
    K^A_\uparrow =& \langle \uparrow \!|_A \otimes  \mathbb{1}_B =& 
        \begin{pmatrix}
            1   &   0   &   0   &   0 \\
            0   &   0  &   1   &   0
        \end{pmatrix} \label{eq:Avup} \\
    K^A_\downarrow =& \langle \downarrow \!|_A \otimes  \mathbb{1}_B =&
        \begin{pmatrix}
            0   &   1   &   0   &   0 \\
            0   &   0   &   0   &   1
        \end{pmatrix}. \label{eq:Avdn}
\end{eqnarray}
Again we can see $K^A_\uparrow$ picks out the rows in a column vector with the first spin up, namely rows 1 and 3, while $K^A_\downarrow$ picks out rows 2 and 4 where the first spin is down.

Using this technology to characterize a degree of freedom may seem like overkill in this case where the Hilbert space is already written in a basis compatible with the decomposition between the degree of freedom integrated out and the degree of freedom that remains. However, it becomes more useful when applied to other, non-local degrees of freedom.

\subsection{Quantum channels for non-local degrees of freedom}

Let us motivate quantum channels for tracing out non-local degrees of freedom by means of duality. Define an action of duality on states in a Hilbert space through a unitary operator $D$,
\begin{equation}
    |\psi\rangle \to |\psi'\rangle = D |\psi\rangle \,,
\end{equation}
with $D^\dagger D = \mathbb{1}$. Duality carries the Hilbert space to another Hilbert space of the same dimension while preserving norms. If the duality is a self-duality, we may think of it as taking a single Hilbert space to itself. In general duality need not square to the identity, though it may. As a unitary operator we can always think of it as a simple quantum channel,
\begin{equation}
    \rho \to \rho' = D \rho D^\dagger \,.
\end{equation}
If we are interested in tracing out non-local degrees of freedom that become local after duality, we can consider a product of quantum channels: first perform duality, then trace out degrees of freedom that are now local. The composition of quantum channels is a single quantum channel, with Kraus operators
\begin{equation}
    \hat{K}_b \equiv K_b D \,.  
\end{equation}
The unitarity of $D$ guarantees these obey
\begin{equation}\label{eq:Dual_Kraus_Relations}
 \sum_b \hat{K}_b^\dagger \hat{K}_b = \mathbb{1}_{AB} \,, \quad \quad \hat{K}_{b'} \hat{K}_b^\dagger = \mathbb{1}_A \delta_{bb'} \,, 
\end{equation}
just like for $K_b$. Thus we can consider the $\hat{K}_b$ to characterize the non-local degrees of freedom $\hat{B}$ mapped to under duality. We can then calculate the entanglement of the non-local degrees of freedom by tracing them out and finding the von Neumann entropy of the non-locally reduced density matrix
\begin{equation}
    \rho_{\hat{A}} = \sum_b \hat{K}_b \rho \hat{K}^\dagger_b \,.
\end{equation}
In general the matrix presentation of $\hat{K}_b$ will not be elementary as the presentation for $K_b$ was.

Again it is useful to consider an example. Return to the system where $A$ and $B$ are each single qubits, and consider the $N=2$ transverse field Ising model with self-dual boundary conditions. The Hamiltonian (\ref{eq:Hamiltonian_self_dual}) is
\begin{equation}
     H_\text{TFIM, self-dual} = -J_z Z_1 Z_2 - J_x X_1 \,.
\end{equation}
The Hamiltonian is exactly self-dual under a composition of Kramer-Wanniers duality and a parity flip of the chain, in this case just exchanging the two dual spins, so we take our duality operation $D$ to be this combination. Then the first dual spin $\hat{A}$ measures the absence or presence of a domain wall for the original spins by the states $+$ and $-$ respectively, while the second dual spin $\hat{B}$ being $+$ or $-$ correlates with the {\it first} original spin being up or down. Thus duality acts on the basis vectors as
\begin{equation}
    D \begin{pmatrix}
        |\!\uparrow \uparrow\rangle \\ |\!\downarrow \uparrow\rangle \\ |\!\uparrow \downarrow\rangle \\ |\!\downarrow \downarrow\rangle 
    \end{pmatrix} = 
    \begin{pmatrix}
        |++\rangle \\   |- - \rangle \\   |-+\rangle \\   |+-\rangle
    \end{pmatrix} \,.
\end{equation}
We can express $D$ as a composition of a permutation $P_2$ and a Hadamard operator acting on each spin,
\begin{equation}
    D = P_2 \cdot H^{\otimes 2} \,,
\end{equation}
with
\begin{equation}
    P_2 = \begin{pmatrix}
        1 & 0 & 0 & 0 \\
        0 & 0 & 0 & 1 \\
        0 & 1 & 0 & 0 \\
        0 & 0 & 1 & 0 
    \end{pmatrix} \,, \quad \quad
    H^{\otimes 2}= \frac{1}{2} \begin{pmatrix}
        1 & 1 & 1 & 1 \\
        1 & -1 & 1 & -1 \\
        1 & 1 & -1 & -1 \\
        1 & -1 & -1 & 1 
    \end{pmatrix} \,,
\end{equation}
giving
\begin{equation}\label{eq:Duality_matrix}
D= \frac{1}{2} \begin{pmatrix}
        1 & 1 & 1 & 1 \\
        1 & -1 & -1 & 1 \\
        1 & -1 & 1 & -1 \\
        1 & 1 & -1 & -1 
    \end{pmatrix} \,.
\end{equation}
This duality transformation is unitary  and squares to the identity $D^\dagger D = D^2 = \mathbb{1}$. Thus it is also hermitian, and since it is real, symmetric as well. Due to this the same matrix $D$ that transforms the basis vectors also acts to transform the components of a given vector.

The Kraus operators $\hat{K}_b \equiv K_b D$ for tracing out the dual $\hat{B}$ degree of freedom are
\begin{eqnarray} \label{eq:Dual_Kraus1}
    \hat{K}_\uparrow &=& 
        \frac{1}{2} \begin{pmatrix}
           1 & 1 & 1 & 1 \\
        1 & -1 & -1 & 1
        \end{pmatrix}  \\
    \hat{K}_\downarrow &=& 
       \frac{1}{2}  \begin{pmatrix}
            1 & -1 & 1 & -1 \\
        1 & 1 & -1 & -1 
        \end{pmatrix}\,.\label{eq:Dual_Kraus2}
\end{eqnarray}
Due to the simple form of the $K_b$ (\ref{eq:vup}), (\ref{eq:vdn}), these are just the top two and bottom two rows of the duality matrix (\ref{eq:Duality_matrix}), respectively.

Before using these results to calculate the entanglement of non-local variables, let us make a general remark about the formalism of Kraus operators to characterize degrees of freedom. Our example here was motivated by duality, which we could think of as another quantum channel. In general, however, any Kraus operators satisfying (\ref{eq:Dual_Kraus_Relations})
\begin{equation}
     \sum_b K_b^\dagger K_b=  \mathbb{1}_{AB} \quad \quad K_{b'} K_b^\dagger = \mathbb{1}_A \delta_{bb'}\,,
\end{equation}
could be treated as characterizing one or more degrees of freedom $B$. Together they imply that the $T_{b'b} \equiv K^\dagger_{b'} K_b$ operators (\ref{eq:Tmatrix}) form a set of orthogonal projectors partitioning Hilbert space by the value of the degree(s) of freedom, as well as a set of raising/lowering operators moving from one subspace to another. Generic such degrees of freedom would be non-local.

\subsection{Measuring non-local entanglement}

We can use our Kraus operators (\ref{eq:Dual_Kraus1}), (\ref{eq:Dual_Kraus2}) representing non-local degrees of freedom to calculate hidden, non-local entanglement. Consider the state
\begin{equation} \label{eq:DW_unentangled_2site}
    |\psi \rangle = |+\!\uparrow\, \rangle \,,
\end{equation}
which is a $\lambda \to 0$ ground state for the self-dual $N=2$ Hamiltonian, as in (\ref{eq:Self_dual_ground_states}). This state has no local entanglement, and indeed tracing out the second spin leaves a pure density matrix. This is obvious by inspection, but we can also do it with the Kraus operators $K_b$,
\begin{eqnarray}
     \rho_A &=& \sum_b K_b\Big( |+\!\uparrow\, \rangle \langle + \! \uparrow\! | \Big)K_b^\dagger\\
     &=&|+ \rangle \langle+| \,. \nonumber
\end{eqnarray}
However the state (\ref{eq:DW_unentangled_2site}) is dual to the entangled state
\begin{equation}
    D |\psi \rangle = \frac{1}{\sqrt{2}} \left(|\!\uparrow \uparrow \rangle + |\!\downarrow \downarrow \rangle \right) \,,
\end{equation}
and thus the non-local entanglement should be present in $|+\uparrow  \rangle$. We can see this by tracing out the non-local degree of freedom $\hat{B}$ using the Kraus operators $\hat{K}_b$ represented as matrices (\ref{eq:Dual_Kraus1}), (\ref{eq:Dual_Kraus2}),
\begin{eqnarray}\label{eq:Nonlocal_Kraus_Ent}
     \rho_{\hat{A}} &=& \sum_b \hat{K}_b\Big( |+\!\uparrow\, \rangle \langle + \! \uparrow\! | \Big)\hat{K}_b^\dagger\\
     &=&\frac{1}{2}\left( |\!\uparrow  \rangle \langle \uparrow  \!| + |\!\downarrow  \rangle \langle \downarrow  \!|  \right)\,, \nonumber
\end{eqnarray}
a mixed state density matrix indicative of entanglement. Notice that the degree of freedom left in (\ref{eq:Nonlocal_Kraus_Ent}) is not the first spin $A$ of (\ref{eq:DW_unentangled_2site}), but the remaining non-local degree of freedom $\hat{A}$ orthogonal to the one traced out.

Thus the state (\ref{eq:DW_unentangled_2site}) is unentangled in terms of the local degrees of freedom $A$, $B$ characterized by Kraus operators $K$, but is entangled in terms of non-local degrees of freedom $\hat{A}$, $\hat{B}$ characterized by the Kraus operators $\hat{K}$. Both sets of degrees of freedom are fundamentally on equal footing, but writing the state as $|+ \uparrow\rangle$ makes the former degrees of freedom manifest. Duality changes which degrees of freedom are manifest, and thus reveals entanglement that was formerly hidden.

Let us make a few remarks about this result.
It is interesting to see how the scrambling of local entanglement occurs thanks to the duality $D = P_2 \cdot H^{\otimes 2}$. The Hadamard factor $H^{\otimes 2}$ contributing to $D$, while complicated-looking, acts as a product on the two degrees of freedom and cannot change local entanglement; the change must be due to the permutation. The Hadamard operator arises because domain walls for the $z$-basis in the original variables appear as $x$-variables in the dual basis. If we undo this for the sake of getting at the essential non-locality, we can consider the modified duality transformation
\begin{equation}
    \tilde{D} \equiv H^{\otimes 2} \cdot D = P_2^T =  \begin{pmatrix}
        1 & 0 & 0 & 0 \\
        0 & 0 & 1 & 0 \\
        0 & 0 & 0 & 1 \\
        0 & 1 & 0 & 0 
    \end{pmatrix}\,.
\end{equation}
The associated Kraus operators for tracing out the non-local $\tilde{B}$ degree of freedom are then
\begin{equation}
    \tilde{K}_\uparrow = 
        \frac{1}{2} \begin{pmatrix}
           1 & 0 & 0 & 0 \\
        0 & 0 & 1 & 0
        \end{pmatrix}\,, \quad \quad 
    \tilde{K}_\downarrow = 
       \frac{1}{2}  \begin{pmatrix}
            0 & 0 & 0 & 1 \\
        0 & 1 & 0 & 0 
        \end{pmatrix}\,.
\end{equation}
The non-local $\tilde{B}$ degree of freedom is not a domain wall, but is the extra ``parity qubit" degree of freedom simply correlated with the spin of the original spin $A$.
We see indeed the first Kraus operator $\tilde{K}_\uparrow$ picks out states with the first spin up, while the Kraus operator $\tilde{K}_\downarrow$ picks out states with the first spin down. To see the domain wall degree of freedom, we could also consider tracing out the non-local $\tilde{A}$ degree of freedom using the Kraus operators $\tilde{K}^A_a = K^A_a \tilde{D}$ with the $K^A_a$ in (\ref{eq:Avup}), (\ref{eq:Avdn}). This gives
\begin{equation}
    \tilde{K}^A_\uparrow = 
        \frac{1}{2} \begin{pmatrix}
           1 & 0 & 0 & 0 \\
        0 & 0 & 0 & 1
        \end{pmatrix}\,, \quad \quad 
    \tilde{K}^A_\downarrow = 
       \frac{1}{2}  \begin{pmatrix}
            0 & 0 & 1 & 0 \\
        0 & 1 & 0 & 0 
        \end{pmatrix}\,,
\end{equation}
and indeed the first picks out the states $|\!\uparrow \uparrow\rangle$, $|\!\downarrow \downarrow\rangle$ with no domain wall, while the second picks out $|\!\uparrow \downarrow\rangle$ and $|\!\downarrow \uparrow\rangle$ where a domain wall is present.

It may seem surprising that something as simple as a permutation can create or remove local entanglement, but for the column vectors the permutation acts on this is indeed the case. The unentangled state $| +\!\!\uparrow \, \rangle = (|\!\!\uparrow \uparrow \rangle+ |\!\!\downarrow \uparrow \rangle)/\sqrt{2}$ is the column vector $(1/\sqrt{2})(1,1,0,0)^T$, while the  duality $\tilde{D} = P_2^T$ turns this into $(1/\sqrt{2})(1,0,0,1)^T = (|\!\!\uparrow \uparrow \rangle+ |\!\!\downarrow \downarrow \rangle)/\sqrt{2}$ which is entangled. A permutation is all that is required to exchange a pair of states with different values of $B$ but the same value of $A$ with a pair of states with different values for both, which is all that is required to generate entanglement. Thus such a ``simple" permutation is in fact quite non-local.

It is also possible to think of entanglement in the language of error-correcting codes. If by ``local errors" we mean operations acting on a single tensor factor of the Hilbert space, then a pair of states cannot make entangled linear combinations if they differ by an error in only one factor, that is if their Hamming distance is 1, because any such linear combination can be written as a product state. But if a pair of states differ by local errors in two or more degrees of freedom (Hamming distance $\geq 2$) then generic linear combinations will be entangled. A non-local operation like duality reshuffles the degrees of freedom and consequently reshuffles which errors are local, turning unentangled states into entangled states and vice versa.

In summary, we have found a way of characterizing degrees of freedom using Kraus operators for the quantum channel that forgets the degree of freedom, and we can use this method to produce the reduced density matrix and thus measure the entanglement. For local degrees of freedom this is trivial, but this method can encode non-local degrees of freedom, such as those obtained through duality, as well. We have thus been able to see how, after a non-local duality, entanglement that disappears from the local variables persists in terms of non-local degrees of freedom.

This formalism using quantum channels is a natural characterization when thinking of the theory in terms of a Hilbert space. It is also possible, and sometimes advantageous, to think about a quantum theory from an algebraic perspective. We turn to this in the next section.

\section{Non-local degrees of freedom from algebraic quantum mechanics}
\label{sec:Algebras}

The quantum channel description of a degree of freedom detailed in Sec.~\ref{sec:QChannels} relied heavily on the Hilbert space formalism of quantum mechanics. The use of partial traces and Kraus operators was possible because the Hilbert space was decomposable into a tensor product of $N$ individual spin degrees of freedom: $\mathcal{H} = \mathbb{C}^{2N} = (\mathbb{C}^2)^{\otimes N}$.  Degrees of freedom are baked into this structure, and we were able to use this to understand how entanglement entropy transformed under duality.

However, this is not the only way to formulate quantum mechanics. It is well understood that algebras, and in particular C$*$-algebras, provide an alternative approach to the Hilbert space formalism. Being replaced by algebras, the Hilbert space is no longer the fundamental object of the theory, but is instead an emergent property. The degree of freedom structure exploited in the previous section can then also be encoded in the defining algebras so that it is passed along to the emergent Hilbert space. Because of this, we will find that algebras provide a natural framework for describing degrees of freedom and how they transform under dualities. We begin with a review of the algebraic approach to quantum mechanics following the works of \citep{Fewster:2019ixc, barata2021pure, Balachandran:2013cq}. We will focus on aspects of algebras and the GNS construction that apply to discrete systems, such as the spin chain of the TFIM. For more information on algebras and entanglement entropy in continuum field theories, see for example \citep{witten_notes_2018,hollands_entanglement_2018}. Those readers familiar with the GNS construction may skip to the second half of Sec.~\ref{sec:Algebras_Rev} where single qubit examples are given. These examples contain details that will be needed for the remainder of the work.

\subsection{A review of algebraic quantum mechanics} \label{sec:Algebras_Rev}
An \textit{algebra} $\A$ is defined as a set of operators, including the identity $\mathbb{1}$, that are closed under two operations:
\begin{enumerate}
    \item Multiplication: if $A,B \in \A$, then the product $AB$ is also an element of $\A$,
    \item Linear combination: if $A,B \in \A$, then the linear combination $\alpha A + \beta B$ is also an element of $\A$, where $\alpha$ and $\beta$ are complex scalars.
\end{enumerate}
Furthermore, a \textit{C$*$-algebra} is an algebra with additional structure, including
\begin{enumerate}
    \item Associativity: $A(BC) = (AB)C$ for all $A,B,C \in \A$, 
    \item Complex conjugation: if $A \in \A$, then the conjugate of $A$ is also an element of the C$*$-algebra, $A^* \in \A$,
    \item Norm: $||\cdot||$ satisfying $||A A^*|| = ||A||^2$ for any element $A$ of $\A$.
\end{enumerate}
The ``$*$'' in ``C$*$'' refers to the inclusion of complex conjugates, while the ``C'' refers to the imposition of a norm. In quantum mechanics, the elements of the algebra correspond to observables, and the $*$ operation is identified as hermitian conjugation. These C$*$-algebras provide the foundation for the algebraic approach to quantum mechanics; any quantum theory may be described by a set of observables forming a C$*$-algebra.

In this formalism the Hilbert space is an emergent property, and the algebra replaces it as the fundamental object of the theory. The Hilbert space may be obtained from the defining algebra using a process called the \textit{GNS construction} \citep{gelfand_imbedding_1943, segal_irreducible_1947}. The construction requires two ingredients: a C$*$-algebra of observables $\A$ and an initial state $\omega$. Here, a state is defined to be a functional over the algebra, assigning real numbers to each element
\begin{equation}
    \omega \, : \, A \rightarrow \mathbb{R}, \quad A \in \A.
\end{equation}
In a more familiar language, states assign real-valued expectation values to observables. While this can be done using a density matrix representation of the state,
\begin{equation} \label{eq:GNS_init_state}
    \omega (A) = \tr ( \rho_\omega A ), \quad A \in \A,
\end{equation}
we are not restricted to representing states as kets or density matrices. 

The GNS construction begins by defining a vector space $\hat\A$ associated with the algebra $\A$. For every element $A$ in the algebra, there is a corresponding vector $|A\rangle$ in $\hat\A$. The state $\omega$ provides the inner product for the vector space,
\begin{equation}
    \langle B | A \rangle = \omega(B^*A)
\end{equation}
where $A$ and $B$ are two elements of $\A$ with associated vectors $|A\rangle$ and $|B\rangle$, respectively. There always exists a vector  $|\mathbb{1}\rangle$ corresponding to the identity operator 
such that
\begin{equation}\label{eq:omegafromI}
    \omega(A) = \langle \mathbb{1} | A \rangle, \quad \omega(\mathbb{1}) = \langle \mathbb{1} | \mathbb{1} \rangle.
\end{equation}
In general, the vector space $\hat\A$ is larger than the physical Hilbert space of the theory. This is because there may exist null states $N$ for which 
\begin{equation}
    \omega(N^*N) = \langle N | N \rangle = 0.
\end{equation}
These states are not physically meaningful and must be removed from the vector space. The space of null states is denoted as
\begin{equation}
    \hat{\mathcal{N}}_\omega = \{ N \, | \, \omega(N^*N) = 0,\, N \in \A \}.
\end{equation}
$\hat{\mathcal{N}}_\omega$ is a subspace of $\hat\A$, and the quotient space $\hat\A / \hat{\mathcal{N}}_\omega$ includes only those vectors with non-zero norm. Vectors in this quotient space then belong to conjugacy classes $|[A]\rangle$ given by all vectors related by an element of $\hat{\mathcal{N}}_\omega$. This quotient space is identified as the Hilbert space,
\begin{equation}
    \mathcal{H}_\omega = \hat\A / \hat{\mathcal{N}}_\omega.
\end{equation}
In addition to the elements of the algebra generating the Hilbert space in this fashion, each element of the algebra is also associated to an operator acting on that Hilbert space. This is done by means of 
 a homomorphism $\pi_\omega$ mapping elements of $\A$ to operator representations on the Hilbert space $\mathcal{H}_\omega$. This map is defined as
\begin{equation} \label{eq:pi_rep}
    \pi_\omega (B) |[A]\rangle = |[BA]\rangle.
\end{equation}
$\pi_\omega(B)$ is the representation of $B$ on $\mathcal{H}_\omega$. Acting with $\pi_\omega(B)$ on the Hilbert space element corresponding to the identity vector $|[\mathbb{1}]\rangle$ gives
\begin{equation}
    \pi_\omega (B) |[\mathbb{1}]\rangle = |[B]\rangle.
\end{equation}
As a result, the homomorphism $\pi_\omega$ can be used to obtain any vector in the Hilbert space by acting with it on $|[\mathbb{1}]\rangle$.

The identity vector plays another important role: it may be used to construct the representation of the initial state's density matrix $\rho_\omega$ by
\begin{equation} \label{eq:rho_identity}
    \rho_\omega = | [\mathbb{1}] \rangle \langle [\mathbb{1}] |\,,
\end{equation}
since it follows from this definition  that
\begin{equation}
    {\rm Tr} \left( \pi_\omega(A) \rho_\omega \right) =  \langle [\mathbb{1}]|[A]\rangle = \omega(A) \,,
\end{equation}
where we used (\ref{eq:omegafromI}).

The GNS construction thus provides a representation of the Hilbert space and operators on that Hilbert space:
\begin{equation}
    \text{GNS} \, : \, (\A, \, \omega) \rightarrow (\mathcal{H}_\omega, \, \pi_\omega).
\end{equation}
Subscripts emphasize that the representation produced by the GNS construction is dependent on the state $\omega$: for a given algebra $\A$, a different choice of initial state can lead to a different Hilbert space.

We now consider two examples of the GNS construction -- one qubit in a pure state and one qubit in a mixed state -- to emphasize details that will be important for our application to dualities.

\subsubsection*{The GNS construction for one qubit in a pure state}
As a first example of the GNS construction, consider constructing an algebra for a single spin. Since the three Pauli matrices, plus the identity, form a basis for all Hermitian operators acting on this spin, the algebra generated by these Pauli operators contains all possible observables. This algebra will be denoted as
\begin{equation} \label{eq:Pauli_gens}
    \A = \langle X, \, Y, \, Z \rangle\,,
\end{equation}
where we have used angle brackets to denote the generators of the algebra. Any element $A$ of this algebra is given by a linear combination of these generators and the identity:
\begin{equation}
    A = \alpha \mathbb{1} + \beta X + i \gamma Y + \delta Z,
\end{equation}
where $\alpha, \, \beta, \, \gamma,$ and $\delta$ are complex scalars. The corresponding vector space is given by $\hat{\A} = \mathbb{C}^4$, with basis vectors $|\mathbb{1}\rangle$, $|X\rangle$, $|Y\rangle$, and $|Z\rangle$.

Before delving into the technicalities of the GNS construction for one qubit, we provide some intuition using the usual ket notion of a state. Take the initial state $\omega$ to be the $+1$ eigenstate of the Pauli $Z$ operator, $|\!\!\uparrow \rangle_z = (1, \, 0)^T$. We may act on this state by any element of the algebra: $X$ flips the state to $|\!\!\downarrow \rangle_z$, $Z$ adds a phase, and $Y$ does both. Taking linear combinations of these actions, we may reach any state in a two-dimensional complex vector space. As such, using the algebra of all Pauli operators and an initial state in the $z$-basis allows us to generate a $\mathbb{C}^2$ Hilbert space.

The full GNS construction is needed to see how the four-dimensional complex vector space $\hat{\A}$ reduces to $\mathbb{C}^2$. Having fixed our algebra $\A$, we also need a state $\omega$. We may think of the state in terms of a density matrix: $\omega(A) = \tr ( \rho A )$.
The most general $\omega$ is given as
\begin{equation}
    \omega(A) = \alpha + \beta \langle X \rangle + i \gamma \langle Y \rangle + \delta \langle Z \rangle\,.
\end{equation}
We then find the null space $\hat{\mathcal{N}}_\omega$ by setting $\omega(N^*N) = 0$,
\begin{eqnarray} 
    0 &=& \tr( \rho N^* N ) \nonumber \\
        &=& |\alpha|^2 + |\beta|^2 + |\gamma|^2 + |\delta|^2 \nonumber \\
        && + \langle X \rangle ( \alpha \beta^* + \beta \alpha^* + \gamma \delta^* + \delta \gamma^* ) \nonumber \\
        && - i \langle Y \rangle ( \alpha \gamma^* - \gamma \alpha^* - \beta \delta^* + \delta \beta^* ) \nonumber \\
        && + \langle Z \rangle ( \alpha \delta^* + \delta \alpha^* - \beta \gamma^* - \gamma \beta^* )\,. \label{eq:full_null_cond}
\end{eqnarray}
Choosing $\omega$ is equivalent to specifying the expectation values of the Pauli operators in (\ref{eq:full_null_cond}). Take, for example, $\langle Z \rangle = 1$ and $\langle X \rangle = \langle Y \rangle = 0$; this corresponds to a pure state pointing in the positive $z$-direction, so we will denote this as $\omega =\:  \uparrow$. In this case (\ref{eq:full_null_cond}) becomes
\begin{equation} \label{eq:Z=1_null_cond}
    0 = |(\alpha + \delta)|^2 + |(\beta - \gamma)|^2.
\end{equation}
Since each modulus square is positive, they must be set to zero independently. Therefore, the null space  corresponds to two constraints,
\begin{equation} \label{eq:null_constraints}
    0 = \alpha + \delta, \quad 0 = \beta - \gamma,
\end{equation}
leaving only two free parameters. The null space $\hat{\mathcal{N}}_\uparrow$ must then be given by $\mathbb{C}^2$, and is spanned by those elements of $\hat\A$ satisfying (\ref{eq:null_constraints}), namely
\begin{equation} \label{eq:null_basis}
    \frac{1}{\sqrt{2}} ( |\mathbb{1}\rangle - |Z\rangle ), \quad \frac{1}{\sqrt{2}} ( |X\rangle + i |Y\rangle ).
\end{equation}
The Hilbert space is found by quotienting $\hat{\A}$ by the null space; here we obtain
\begin{equation}
    \mathcal{H}_\uparrow = \hat{\A} / \hat{\mathcal{N}}_\uparrow = \mathbb{C}^2,
\end{equation}
as expected for a single qubit. The equivalence classes acting as a basis for the Hilbert space can then be defined as the vectors orthogonal to (\ref{eq:null_basis}), plus arbitrary null vectors,
\begin{eqnarray} \label{eq:hilbert_basis}
    |\!\uparrow\,\rangle &\equiv& \frac{1}{\sqrt{2}} ( |\mathbb{1}\rangle + |Z\rangle ) + {\rm null}\,, \\
    |\!\downarrow\,\rangle &\equiv& \frac{1}{\sqrt{2}} ( |X\rangle - i |Y\rangle ) + {\rm null}\,.\label{eq:hilbert_basis2}
\end{eqnarray}
To see that the names given for the vectors  (\ref{eq:hilbert_basis}, \ref{eq:hilbert_basis2}) are appropriate, we construct the representation $\pi_\omega$ of the algebra $\A$ on the Hilbert space $\mathcal{H}_\omega$. The representation $\pi_{\uparrow}(X)$ is defined using (\ref{eq:pi_rep}) and the usual Pauli product operations:
\begin{eqnarray}
    \pi_\uparrow (X) |\!\uparrow\rangle &=& \frac{1}{\sqrt{2}} \left(|X \mathbb{1}\rangle + |XZ\rangle \right) = |\!\downarrow\rangle \,,\\
    \pi_\uparrow (X) |\!\downarrow\rangle &=& \frac{1}{\sqrt{2}} \left(|XX \rangle -i |XY\rangle \right) = |\!\uparrow\rangle\,.
\end{eqnarray}
Any null states are carried into other null states but do not affect the equivalence class.
If we express the basis vectors $|\!\uparrow\,\rangle$ and $|\!\downarrow\,\rangle$ as the two-component vectors $(1,0)^T$ and $(0,1)^T$ respectively, the representation of $X$ is explicitly given as
\begin{equation}
    \pi_\uparrow(X) = 
        \begin{pmatrix}
            0   &   1 \\
            1   &   0
        \end{pmatrix}.
\end{equation}
Similarly, the remaining two Pauli operators are found to be 
\begin{equation}
    \pi_\uparrow(Y) = 
        \begin{pmatrix}
            0   &   -i \\
            i   &   0
        \end{pmatrix}, \quad
    \pi_\uparrow(Z) = 
        \begin{pmatrix}
            1   &   0 \\
            0   &   -1
        \end{pmatrix}.
\end{equation}
Thus the observables $X$, $Y$, and $Z$ emerge correctly from the representation $\pi_\uparrow$, and the GNS construction beginning with the state $\omega = \: \uparrow$ generates the entire one-qubit Hilbert space and its observables.

From (\ref{eq:rho_identity}) we expect that the density matrix for the state $\omega = \uparrow$ used in the GNS construction can be built from the equivalence class of the identity operator. Indeed we see that $|\mathbb{1}\rangle$ is in the equivalence class of $|\!\uparrow\rangle$,
\begin{equation}\label{eq:Identity_up}
   | \mathbb{1} \rangle = \frac{1}{2} \big( |\mathbb{1}\rangle + |Z\rangle \big) +  \frac{1}{2} \big( |\mathbb{1}\rangle - |Z\rangle \big) \,,
\end{equation}
with the second term being null.

The construction is similar for any initial state given by a pure state of a qubit. Pure states may be represented as unit vectors on a Bloch sphere, related to $\omega = \uparrow$ by a rotation. The GNS construction for another pure state proceeds equivalently to the above, giving $\mathcal{H}_\omega = \mathbb{C}^2$, but with the representation of the algebra $\pi_\omega(\A)$ emerging as a rotation of the Pauli matrices. The construction using a mixed initial state is different, and we turn to it now.

\subsubsection*{The GNS construction for one qubit in a mixed state}
The GNS construction is capable of recognizing the difference between a pure state and a mixed state; this property will prove essential for understanding the transformation of entanglement entropy in the algebraic formalism. If the initial state $\omega$ is mixed, such that $\tr(\rho_\omega^2) < 1$, the representation $\pi_\omega$ given by the GNS construction will be \textit{reducible} \citep{Balachandran:2013cq, balachandran_entanglement_2013}. The reducibility of a representation may be determined by looking for subspaces of the Hilbert space invariant under transformations of $\pi_\omega(\A)$. 

To see this, consider two orthogonal states $|[B]\rangle$ and $|[C]\rangle$ in $\mathcal{H}_\omega$. If there is no element $A\in\A$ that can be used to transform $|[B]\rangle$ into $|[C]\rangle$ (or vice-versa),
\begin{equation} \label{eq:inv_subspaces}
    \langle[C]| \pi_\omega(A) |[B]\rangle = 0,
\end{equation}
then $|[B]\rangle$ and $|[C]\rangle$ belong to separate invariant subspaces of $\mathcal{H}_\omega$. The Hilbert space is then reducible and $\pi_\omega$ can be written as a direct sum over irreducible representations acting on the invariant subspaces of $\mathcal{H}_\omega$. The identity vector $|\mathbb{1}\rangle$ can then be decomposed on these invariant subspaces (here labeled by $i$),
\begin{equation}\label{eq:Identity_decomp}
    |\mathbb{1}\rangle = \sum_i |\mathbb{1}_i\rangle.
\end{equation}
Equation (\ref{eq:inv_subspaces}) applies to the components of $|\mathbb{1}\rangle$ as well,
\begin{equation}
    \langle\mathbb{1}_i| \pi_\omega(A) |\mathbb{1}_j\rangle = 0, \quad i \neq j.
\end{equation}
Expressing the initial state's density matrix $\rho_\omega$ in terms of the identity vector as in (\ref{eq:rho_identity}), we find
\begin{eqnarray}\label{eq:rho_reducible}
    \omega(A) &=& \langle \mathbb{1} | \pi_\omega (A) | \mathbb{1} \rangle 
        = \sum_{i,j} \langle \mathbb{1}_i | \pi_\omega (A) | \mathbb{1}_j \rangle
        = \tr \left( \pi_\omega (A) \sum_{i,j} | \mathbb{1}_j \rangle \langle \mathbb{1}_i | \right) \nonumber \\
        &=& \tr \left( \pi_\omega (A) \sum_i | \mathbb{1}_i \rangle \langle \mathbb{1}_i | \right) \,,
\end{eqnarray}
where in the final line the terms $| \mathbb{1}_j \rangle \langle \mathbb{1}_i | $ with $j \neq i$ disappeared because there is no $\pi_\omega(A)$ for which they do not vanish.
This final line implies the initial state density matrix is itself reducible, and can be expressed as a weighted sum over smaller density matrices,
\begin{equation} \label{eq:rho_mixed}
    \rho_\omega = \sum_i |\mathbb{1}_i \rangle \langle \mathbb{1}_i| = \sum_i \lambda_i \rho_i\,,
\end{equation}
where the $\lambda_i$ are the norms of the individual  $|\mathbb{1}_i\rangle$, and the $\rho_i$ are normalized density matrices,
\begin{equation}
    \lambda_i \equiv \langle \mathbb{1}_i | \mathbb{1}_i \rangle  \,, \quad \quad \rho_i \equiv  \frac{1}{\lambda_i}|\mathbb{1}_i \rangle \langle \mathbb{1}_i| \,.
\end{equation}
 Therefore, if a representation produced by the GNS construction is reducible, the initial state was a mixed state.

In the example of $\omega  = \uparrow$, there were no invariant subspaces in the Hilbert space; any orthogonal linear combinations of 
 the two basis vectors $|\!\uparrow\,\rangle$ and $|\!\downarrow\,\rangle$ defined in (\ref{eq:hilbert_basis}) could be transformed into each other using $\pi_\uparrow(X)$, $\pi_\uparrow(Y)$, and $\pi_\uparrow(Z)$. The representation $\pi_\uparrow$ was then irreducible, indicating that the initial state was pure. Indeed, we began the example with a pure state.

Consider instead beginning the GNS construction for a single qubit in a mixed state,
\begin{equation}
    \rho = \frac{1}{2}
        \begin{pmatrix}
            1   &   0 \\
            0   &   1
        \end{pmatrix} \,.
\end{equation}
We denote this state $\omega = \mathbb{1}/2$. We use the same algebra, so all of the steps in the construction up through the general null space condition (\ref{eq:full_null_cond}) are the same. The difference is in the expectation values provided by the new state:
\begin{equation}
    \langle X \rangle = \langle Y \rangle = \langle Z \rangle = 0.
\end{equation}
The null space condition (\ref{eq:full_null_cond}) then becomes
\begin{equation}
    0 = |\alpha|^2 + |\beta|^2 + |\gamma|^2 + |\delta|^2.
\end{equation}
Each term must go to zero separately, leaving no free parameters to define the null space. The null space is then trivial, $\hat{\mathcal{N}}_{\mathbb{1}/2} = \emptyset$, and the Hilbert space is
\begin{equation}
    \mathcal{H}_{\mathbb{1}/2} = \hat{\A} / \hat{\mathcal{N}}_{\mathbb{1}/2} = \mathbb{C}^4.
\end{equation}
This Hilbert space is twice as large as the previous example. A convenient basis of states is 
\begin{equation}\label{eq:Reducible_basis}
\begin{gathered}
    |\!\uparrow\rangle \equiv \frac{1}{\sqrt{2}} ( |\mathbb{1}\rangle + |Z\rangle ), \quad |\!\downarrow\rangle \equiv \frac{1}{\sqrt{2}} ( |X\rangle - i |Y\rangle ),  \\
    |\!\uparrow'\rangle \equiv \frac{1}{\sqrt{2}} ( |X\rangle + i |Y\rangle ), \quad |\!\downarrow'\rangle \equiv \frac{1}{\sqrt{2}} ( |\mathbb{1}\rangle - |Z\rangle )\,,
\end{gathered}
\end{equation}
and calculating the representation $\pi_{\mathbb{1}/2}$ we find that $X$, $Y$ and $Z$ act on $|\!\uparrow\rangle$, $ |\!\downarrow\rangle$ as one qubit and $|\!\uparrow'\rangle$, $ |\!\downarrow'\rangle$ as another, but never mix the two sets of states; they are thus invariant subspaces. (The action of the operators is the same as in the pure state case, except there $|\!\uparrow'\rangle$, $ |\!\downarrow'\rangle$ were null states, while here they are physical.) Since the Hilbert space splits into two invariant subspaces,
\begin{equation}
    \mathcal{H}_{\mathbb{1}/2} = \mathbb{C}^2 \oplus \mathbb{C}^2,
\end{equation}
the representation $\pi_{\mathbb{1}/2}$ is reducible. The GNS construction has thus verified that the initial state was indeed mixed.

We expect the identity state $|\mathbb{1}\rangle$ to decompose on the two subspaces as in (\ref{eq:Identity_decomp}), and indeed we have
\begin{eqnarray}
    | \mathbb{1} \rangle &=& \frac{1}{2} \big( |\mathbb{1}\rangle + |Z\rangle \big) +  \frac{1}{2} \big( |\mathbb{1}\rangle - |Z\rangle \big) \\
    &=& \frac{1}{\sqrt{2}} \big( |\!\uparrow\rangle  +  |\!\downarrow'\rangle \big) \,, \nonumber
\end{eqnarray}
the first line of which is identical to (\ref{eq:Identity_up}), but is now interpreted differently as the second vector is no longer null.
The density matrix is
\begin{eqnarray}
    \rho_{\mathbb{1}/2} = |\mathbb{1}\rangle\langle \mathbb{1}| &=& \frac{1}{2} \Big( |\!\uparrow\rangle\langle\uparrow\!| + |\!\downarrow'\rangle\langle\downarrow'\!| +|\!\uparrow\rangle\langle\downarrow'\!| + |\!\downarrow'\rangle\langle\uparrow\!| \Big)\\
     &=& \frac{1}{2} \Big( |\!\uparrow\rangle\langle\uparrow\!| + |\!\downarrow'\rangle\langle\downarrow'\!|  \Big) \,,\nonumber
\end{eqnarray}
where in the second line we dropped $|\!\uparrow\rangle\langle\downarrow'\!|$ and  $|\!\downarrow'\rangle\langle\uparrow\!|$ since there is no observable $A$ that connects the two sectors, as in (\ref{eq:rho_reducible}). Thus if we imagine purifying our system, one of our qubits $\{ |\!\uparrow\rangle, |\!\downarrow\rangle\}$ represents the degree of freedom entangled with one state of the environment, while the second qubit $\{ |\!\uparrow'\rangle, |\!\downarrow'\rangle\}$ represents the degree of freedom entangled with an orthogonal state of the environment.

Our starting state $\omega = \mathbb{1}/2$ was maximally mixed and treats $X$, $Y$, and $Z$ all on the same footing. The basis (\ref{eq:Reducible_basis}) appears to single out $|Z\rangle$, but one can equally well define bases singling out $|X\rangle$, $|Y\rangle$, or any combination on the Bloch sphere, which also span the space $\mathbb{C}^2 \oplus \mathbb{C}^2$.

\subsection{Characterizing degrees of freedom via subalgebras} \label{sec:Algebras_DoFs}
Thus far our examples have involved a single qubit degree of freedom; the algebra for one qubit is
\begin{equation}
    \A_1 = \langle X,\, Y,\, Z \rangle.
\end{equation}
Due to the relation $ZX=iY$, we can characterize the same algebra using just the generators $X$ and $Z$,
\begin{equation} \label{eq:min_gen_A}
    \A_1 = \langle X,\, Z \rangle \,.
\end{equation}
In principle another pair $Y, Z$ or $Z, X$ generates the algebra just as well, but on the level of the algebra alone, without considering a particular representation, all of these pairs have identical algebraic relations and thus are equivalent. Which one is diagonal in the natural basis produced by the GNS construction (the basis including $|[\mathbb{1}]\rangle$ and orthogonal vectors) depends on the initial state $\omega$ chosen. For $\omega = \uparrow$ the representation of $Z$ on the Hilbert space is diagonal; its eigenstates make up the natural basis coming from the GNS construction and we can call it a ``basis" operator. For this same initial state the operator $X$ exchanges the two basis vectors, which we can term a ``swap" operator. Since a single Pauli operator is not enough to generate the algebra $\A_1$, we can think of a single degree of freedom being characterized by both a basis operator $Z$ and a swap operator $X$.

Let us compare this algebraic characterization of a degree of freedom to the quantum channel perspective. In the latter a degree of freedom is represented by a set of Kraus operators $K_b$. We noted from the Kraus operators we could define the matrix $T_{ab} \equiv K^\dagger_a K_b$  (\ref{eq:Tmatrix}). Swap operators between degrees of freedom are represented as $T_{ab} + T_{ba}$, while basis operators are generated as linear combinations of the diagonal elements, $\sum_b b T_{bb}$, with the coefficient $b$ an eigenvalue characterizing the basis vector. Thus the quantum channel perspective also leads to basis operators and swap operators.

This algebraic description of a single qubit can be readily generalized to an algebra for $N$ qubits. We consider the algebra 
\begin{equation}
    \A_N = \langle X_1,\, Z_1;\, X_2,\, Z_2;\, \dots;\, X_N,\, Z_N \rangle \,,
\end{equation}
where each $X_i, Z_i$ obey the usual Pauli algebra $X_i^2 = Z_i^2 = \mathbb{1}$, $Z_i X_i = -X_i Z_i = i Y_i$, and $\{X_i, Z_i\}$ commute with $\{X_j, Z_j\}$ for $i \neq j$; semicolons have been used to visually separate generators associated to each qubit. A simple construction of the Hilbert space proceeds from choosing the initial state $\omega$ to be a pure, separable state, for which the expectation value of a product of Pauli operators equals the product of the individual expectation values; further taking $\langle Z_i \rangle = 1$ and $\langle X_i \rangle= \langle Y_i \rangle = 0$ gives the familiar representation with each $Z_i$ a basis operator and each $X_i$ a swap operator.

The generators $X_i, Z_i$ characterizing a single degree of freedom form a subalgebra of the total algebra $\A_N$. We may implement the partial trace of a degree of freedom by removing the generators of this subalgebra \citep{Balachandran:2013cq, balachandran_entanglement_2013}. For example, to trace out the $i^\text{th}$ spin, we remove $X_i, Z_i$ from the list of generators and regenerate an algebra for $N-1$ spins,
\begin{equation}
    \A_{N-1} = \langle X_1,\, Z_1;\, \dots;\, X_{i-1},\, Z_{i-1};\; X_{i+1},\, Z_{i+1};\, \dots;\, X_N,\, Z_N \rangle.
\end{equation}
We must also describe what happens to the GNS state $\omega_N$. Tracing out a degree of freedom induces a reduced state $\omega_{N-1}$ inheriting its values from $\omega_N$,
\begin{equation}
    \omega_{N-1}(A) = \omega_N(A) \,, \quad \forall \, A \in \A_{N-1} \,.
\end{equation}
In general we could trace out multiple degrees of freedom, which characterize a larger subalgebra, in an analogous fashion.

We now have the technology to calculate entanglement entropy in the algebraic framework \citep{Balachandran:2013cq, balachandran_entanglement_2013}. Suppose we wish to determine the entanglement entropy of a pure state $|\psi\rangle_{AB} \in \mathcal{H}_A \otimes \mathcal{H}_B$.\footnote{Here $A$ and $B$ are subsystems, not elements of an algebra. From here on, we will use $O$ to denote generic algebra elements to avoid confusion.} In the algebraic framework this means we have an algebra $\A_{AB}$ and an initial state $\omega_{AB}$ that through the GNS construction produces a representation $\pi_{\omega_{AB}}$ on the Hilbert space that is irreducible. After obtaining the reduced algebra $\A_A$ by removing the generators of the subalgebra $\A_B$, we define a restricted state $\omega_A$ acting on $\A_A$ by requiring that it agree with the state $\omega_{AB}$ for any operator in $\A_A$:
\begin{equation}
    \omega_A (O) = \omega_{AB}(O), \quad \forall \, O \in \A_A.
\end{equation}
Performing the GNS construction using $\A_A$ and the restricted state $\omega_A$ gives a representation $\pi_{\omega_A}$. If $\pi_{\omega_A}$ is reducible, the lessons of Sec.~\ref{sec:Algebras_Rev} indicate that the restricted state $\omega_A$ is mixed. The subsystems $A$ and $B$ must then be entangled. The reduced density matrix $\rho_A$ is then given as a sum over invariant subspaces as in (\ref{eq:rho_mixed}). The normalizations $\lambda_i = \langle \mathbb{1}_i| \mathbb{1}_i \rangle$ then can be used directly in the von Neumann formula to calculate the entanglement entropy of $A$ in the state $\omega_{AB}$,
\begin{equation}
    S(A) = - \sum_i \lambda_i \log \lambda_i.
\end{equation}
To illustrate this process, let us calculate the  entanglement entropy of the state (\ref{eq:DW_unentangled_2site}), repeated here for convenience,
\begin{equation} \label{eq:no_ent_example}
    |\psi\rangle = |+\!\uparrow\, \rangle\,,
\end{equation}
from the algebraic perspective. It is clear that this state has no local entanglement, so we ought to be able to verify this. The algebra is generated by the basis and swap operators for both spins,
\begin{equation} \label{eq:N=2_max_algebra}
    \A_{AB} = \langle X_1, \, Z_1 ;\, X_2, \, Z_2 \rangle\,,
\end{equation}
where $X_1, Z_1$ generate the degree of freedom $A$ and $X_2$, $Z_2$ generate the degree of freedom $B$. The GNS state $\omega_{AB}$ encodes the state (\ref{eq:no_ent_example}) as
\begin{equation} \label{eq:N=2_omega}
\begin{gathered}
    \omega_{AB}(\mathbb{1}) =    \omega_{AB}(X_1) = \omega_{AB}(Z_2) = 1 \,,  \\
    \omega_{AB}(X_1 Z_1) = \omega_{AB}(Z_1) = \omega_{AB}(X_2) = \omega_{AB}(X_2 Z_2) = 0 \,,
\end{gathered}
\end{equation}
with the expectation values of products of $1$ and $2$ Pauli elements given as the product of individual expectation values.

Choosing to trace out the second spin $B$, we restrict to a subalgebra of the first,
\begin{equation}
    \A_A = \langle X_1,\, Z_1 \rangle, \quad O = \alpha \mathbb{1} + \beta X_1 + \gamma Z_1 + \delta X_1 Z_1 \,\, \in \,\, \A_A,
\end{equation}
where $\alpha$, $\beta$, $\gamma$, and $\delta$ are complex numbers. The state (\ref{eq:N=2_omega}) is used to find the state $\omega_A$ restricted to this subalgebra,
\begin{equation}
    \omega_A(\mathbb{1}) = \omega_A(X_1) = 1, \quad \omega_A(Z_1) = \omega_A(X_1 Z_1) = 0.
\end{equation}
Using this reduced state to find vectors with zero norm,
\begin{equation}
    0 = \omega_A(N^* N) = |(\alpha + \beta)|^2 + |(\gamma - \delta)|^2,
\end{equation}
the null space is found to be $\mathbb{C}^2$ and spanned by the vectors
\begin{equation}
    \frac{1}{\sqrt{2}} ( |\mathbb{1}\rangle - |X_1\rangle ),  \quad \frac{1}{\sqrt{2}} ( |Z_1\rangle + |X_1 Z_1\rangle ).
\end{equation}
The Hilbert space is then spanned by the remaining two vectors orthogonal to these, namely
\begin{equation}
    |+\rangle \equiv \frac{1}{\sqrt{2}} ( |\mathbb{1}\rangle + |X_1\rangle ),  \quad |-\rangle \equiv \frac{1}{\sqrt{2}} ( |Z_1\rangle - |X_1 Z_1\rangle )\,,
\end{equation}
plus arbitrary null vectors.
There is no invariant subspace, with $\pi_{\omega_A}(Z_1)$ acting as a swap operator. Therefore the representation is irreducible, and the state (\ref{eq:no_ent_example}) has no local entanglement, matching our expectation.

\subsection{Duality and non-local degrees of freedom}

We now use our algebraic tools to investigate the transformation of entanglement entropy under duality. Consider again the case of two qubits, with algebra given in terms of basis and swap operator generators as in (\ref{eq:N=2_max_algebra}). Kramers-Wannier duality naturally acts on the algebra, as in (\ref{eq:dual_tauX}), (\ref{eq:dual_tauZ}). To match the duality of the quantum channel approach, let us compose KW duality with a parity flip on the spins; we also change the labels from $1/2$, $3/2$ to $1$, $2$ and write $X$, $Z$ instead of $\mathcal{X}$, $\mathcal{Z}$ to emphasize this is a self-duality. Duality then acts as
\begin{equation}
    X_1 \to Z_1 Z_2 \,, \quad Z_1 \to X_2 \,, \quad \quad X_2 \to Z_1 \,, \quad Z_2 \to X_1 X_2\,.
\end{equation}
We can thus equivalently characterize the algebra as
\begin{equation} \label{eq:TFIM_algebra_nonlocal}
    \A_{AB} = \langle X_2, \, Z_1 Z_2; \, X_1 X_2, \, Z_1 \rangle \,.
\end{equation}
We emphasize that the algebra has not changed; we are using a different set of four generators but the same algebra is generated. However, duality has expressed the same algebra in terms of different subalgebras,
the first generated by $X_1$ and $Z_1 Z_2$, which we may call $\hat{A}$, and the second by $X_1 X_2$ and $Z_2$, which we may call $\hat{B}$, each corresponding to a degree of freedom distinct from the the ones associated to $X_1$, $Z_1$ and $X_2$, $Z_2$; thus $\A_{\hat{A}\hat{B}} = \A_{AB}$. That these degrees of freedom are non-local in terms of the original ones is explicit. Tracing out by one of the new, non-local degrees of freedom by removing the appropriate generators leaves behind a subalgebra associated to the other. 

Consider again the state  $|\psi\rangle = |+\!\uparrow\, \rangle$, encoded in the GNS state $\omega$ in (\ref{eq:N=2_omega}). $|\psi\rangle$ has no local entanglement, but since it is dual to the entangled state $(|\!\uparrow\uparrow\,\rangle + |\!\downarrow\downarrow\,\rangle)/\sqrt{2}$, we expect that $|\psi\rangle$ is hiding non-local entanglement. Let us use the algebraic formulation to verify this.
The entanglement entropy between the non-local degrees of freedom can be found by tracing out $\hat{B}$ and leaving $\hat{A}$,
\begin{equation}
    \A_{\hat{A}} = \langle X_2 ,\, Z_1 Z_2 \rangle, \quad O = \alpha \mathbb{1} + \beta X_2 + \gamma Z_1 Z_2 + \delta X_2 Z_1 Z_2 \,\, \in \,\, \A_{\hat{A}} \,.
\end{equation}
Restricting $\omega$ (\ref{eq:N=2_omega}) to this subalgebra gives
\begin{equation}
    \omega_{\hat{A}}(\mathbb{1}) = 1, \quad \omega_{\hat{A}}(X_2) = \omega_{\hat{A}}(Z_1 Z_2) = \omega_{\hat{A}}(X_2 Z_1 Z_2) = 0 \,.
\end{equation}
We see the state now has the apparance of being maximally mixed, since all nontrivial expectation values vanish.
The null space is empty since
\begin{equation}
    0 = \omega_{\hat{A}}(N^* N) = |\alpha|^2 + |\beta|^2 + |\gamma|^2 + |\delta|^2 \,,
\end{equation}
constrains all parameters to be zero. The Hilbert space is then spanned by all four vectors $\{|\mathbb{1}\rangle,\, |X_2\rangle,\, |Z_1 Z_2\rangle,\, |X_2 Z_1 Z_2\rangle \}$. There are two invariant subspaces defined by the vectors
\begin{equation}
    \left\{ |+\rangle \equiv \frac{1}{\sqrt{2}} ( |\mathbb{1}\rangle + |X_2\rangle ), \quad |-\rangle \equiv \frac{1}{\sqrt{2}} ( |Z_1 Z_2\rangle - |X_2 Z_1 Z_2\rangle ) \right\}
\end{equation}
\begin{equation}
    \left\{ |-'\rangle \equiv \frac{1}{\sqrt{2}} ( |\mathbb{1}\rangle - |X_2\rangle ), \quad |+'\rangle \equiv \frac{1}{\sqrt{2}} ( |Z_1 Z_2\rangle + |X_2 Z_1 Z_2\rangle ) \right\} \,.
\end{equation}
$\pi_{\omega_{\hat{A}}}(Z_1Z_2)$ acts as a swap operator relating $|+\rangle$ to $|-\rangle$, and $|+'\rangle$ to $|-'\rangle$, but no operator relates the primed and unprimed states.
The representation is then reducible and the state is entangled. 

The identity vector $|\mathbb{1}\rangle$ decomposes on these invariant subspaces as $(|+\rangle + |-'\rangle)/\sqrt{2}$. The entanglement entropy is then found by calculating the norms of these contributions:
\begin{equation}
    \langle + | + \rangle = \langle -' | -' \rangle = \frac{1}{2}
\end{equation}
\begin{equation}
    \implies S(a) = - \langle + | + \rangle \log \langle + | + \rangle - \langle -' | -' \rangle \log \langle -' | -' \rangle = \log 2.
\end{equation}
While there is no entanglement between the local degrees of freedom in state $|+ \uparrow \rangle$, the non-local degrees of freedom given by the swap and basis operators in (\ref{eq:TFIM_algebra_nonlocal}) are indeed entangled. We again find that entanglement has not been lost, but has been transferred to non-local degrees of freedom provided by the Kramers-Wannier duality.

\section{Conclusions}
\label{sec:Conclusions}

By characterizing local and non-local degrees of freedom, we have studied the transformation of entanglement entropy under duality. Implementing a simple, self-dual version of Kramers-Wannier duality as an example, we found that local entanglement is not generally preserved by duality transformations. Instead, entanglement (or lack thereof) is transferred to non-local degrees of freedom. These non-local degrees of freedom were described in two ways: first by quantum channels using Kraus operators implementing the partial trace, and second in the algebraic formulation of quantum mechanics as subalgebras. The two approaches may be used independently to show the transformation of entanglement and degrees of freedom under dualities -- we provide both to give a more complete understanding.

Recent work by \citep{Radicevic:2016tlt} has also explored the transformation of entanglement entropy under dualities in discrete models. There it was found that the Kramers-Wannier duality preserves local entanglement. Instead of shifting entanglement to non-local degrees of freedom, it is moved among the local degrees of freedom. These results were obtained using an algebraic approach that differs slightly from our own. The work of \citep{Radicevic:2016tlt} restricts the algebra of observables to a subalgebra generated only by $Z_i$ and $X_i X_{i+1}$, taking density matrices to be elements of this subalgebra. In the present work, we consider algebras containing all possible observables (generated by swap and basis operators) so that all initial pure states give the same Hilbert space under the GNS construction. In doing so, we take dualities to behave like global symmetries which, unlike gauge symmetries, do not restrict the algebra of observables. (See \citep{DeHaro:2018yqx, deharo2019theoretical} for more discussion on the nature of dualities.)

Kramers-Wannier duality may be implemented in other ways not considered in this work. For example, \citep{radicevic_spin_2019} implements self-dual boundary conditions for the TFIM on a periodic chain, which requires a projection on the state space. (This is in contrast to the periodic boundary conditions mentioned in sec.~2.1.2, which have no projection but are not self-dual.)
In \citep{radicevic_spin_2019}, only dual states with an even number of domain walls are kept when dualizing the theory. Odd numbers of domain walls would correspond to an anti-periodic chain, and these states are projected out. One must then project out half of the spin states to match the reduced domain wall state space. The true duality is then between theories of spins mod $\mathbb{Z}_2$. In using self-dual boundary conditions on an open chain, we implement the duality without projection, as we see explicitly in the matching of the states and their energies on either side of our duality.

The relationship between entanglement and duality has also been explored in continuum quantum field theories. For example, the work of \citep{headrick_bosefermi_2013} considers the transformation of entanglement entropy under bosonization, a duality between a compact boson and a free fermionic theory. There it was found that after the fermionic theory has been properly gauged, the entire Renyi entropy spectrum agrees between the two theories. This furthered the claim that entanglement entropies should agree for equivalent theories, in tension with our present results. This difference may be due to the discrete nature of the Kramers-Wannier duality. Since bosonization is a duality between two continuum field theories, locality is determined by the geometry of spacetime. There is no sense of a ``dual spacetime'' on which the dual theory is defined -- both theories share the same definition of locality, an important ingredient in quantum field theory. This is not true for the Kramers-Wannier duality, where the spin chain is discrete and a notion of locatity comes only from the Hamiltonian. Duality moves the chain to a dual lattice, changing the definition of locality. It would be interesting to consider the fate of our results in the continuum limit.\footnote{We thank Matthew Headrick for useful discussions on this point.} We leave these questions for future work.

There are several areas of research that rely on a concrete understanding of non-local degrees of freedom for which our characterizations using Kraus operators and algebras could be useful. As they currently stand, these tools are most applicable to discrete systems such as the TFIM. For example, the pursuit of a discrete holographic model (see for instance \citep{gesteau_holographic_2022,jahn_boundary_2022,basteiro_towards_2022} and references therein for recent work) could be a natural candidate for the methods detailed here. Beyond such discrete models, a generalization to continuous degrees of freedom would be necessary. Such a framework could shed light on topics such as black hole complementarity \citep{susskind_stretched_1993}, where it is expected that degrees of freedom behind the horizon are related to highly non-local degrees of freedom accessible to an outside observer. We hope that the tools developed here will assist in the pursuit of these and other important questions.

\section*{Acknowledgments}

We thank Graeme Smith and Matthew Headrick for useful discussions. The authors are supported by the Department of Energy under grant DE-SC0010005.

\end{document}